\documentclass{elsart}
\usepackage{graphicx}
\usepackage{amssymb}
\begin{document}
\begin{frontmatter}

\title{Critical properties of random anisotropy magnets}

 \author[label1,label2]{M.\ Dudka\corauthref{cor1}}\ead{maxdudka@icmp.lviv.ua},
 \author[label2]{R.\ Folk}\ead{folk@tphys.uni-linz.ac.at},
 \author[label3,label2]{Yu.\ Holovatch}
 \ead{hol@icmp.lviv.ua}
 \corauth[cor1]{Corresponding author.}
 \address[label1]
{Institute for Condensed Matter Physics of the National Academy of
Sciences of Ukraine, UA-79011 Lviv, Ukraine}
\address[label2]{Institut f\"ur Theoretische Physik, Johannes Kepler
 Universit\"at Linz, A-4040 Linz, Austria}
 \address[label3]{Institute for Condensed Matter Physics
 and Ivan Franko National University of Lviv,
                  UA-79011 Lviv, Ukraine}
\begin{abstract}
The  problem of critical behaviour of three dimensional random
anisotropy magnets, which constitute a wide class of disordered
magnets is considered. Previous results obtained in experiments,
by Monte Carlo simulations and within different theoretical
approaches give evidence for a second order phase transition for
anisotropic distributions of the local anisotropy axes, while for
the case of isotropic distribution such transition is absent. This
outcome is described by renormalization group in its field
theoretical variant on the basis of the random anisotropy model.
Considerable attention is paid to the investigation of the
effective critical behaviour which explains the observation of
different behaviour in the same universality class.
\end{abstract}

\begin{keyword}
disordered magnets  \sep random anisotropy
 \sep renormalization group
\PACS 61.43.-j  \sep  64.60.Ak
\end{keyword}
\end{frontmatter}


\section{Introduction}\label{I}

The influence of  structural disorder on the critical behaviour
remains to be one of the most attractive problems of  phase
transitions theory \cite{Holovatch_order}. It is known that  even
small disarrangement in the structure of ideal physical systems
may have  crucial consequences for their behaviour near a critical
point \cite{Pelissetto02,Folk03}. In particular,   weak quenched
disorder in  magnetic systems
 can not only change the characteristics  of the second
order phase transition or the order of the phase transition into
the low-temperature phase,  it can modify also the nature of this
phase, producing spin-glass order. Among disordered magnets
depending on the type of randomness of their structure random site
\cite{Pelissetto02,Folk03}, random field \cite{Belanger91}, and
random anisotropy \cite{Cochraine78} systems can be discriminated.
All these magnets were extensively investigated, however, in our
view, less attention was paid to the critical behaviour of the
last class of systems i.e. random anisotropy magnets, in
particular to random anisotropy effects on  second order phase
transitions. This motivates us to present an analysis of the
problem by the field-theoretical renormalization group (RG)
methods \cite{rgbooks}, which are standard for studying the second
order phase transition of ideal systems;  we also briefly review
the present state of the investigations of the low-temperature
phase of the random anisotropy magnets. The main goal of our study
is to emphasize the crucial role played by  distribution of the
random axes in defining the university class of random anisotropy
magnets, and hence the very possibility of a second order phase
transition and its scenario. Moreover, in our study we will
address the effective behaviour of random anisotropy magnets: it
is frequently encountered in experimental and MC studies but less
attention has been paid to it in the theoretical analysis.

The paper is organized as follows: in the next Section we give a
review  of experimental, numerical and theoretical analysis of the
low-temperature phase in random anisotropy magnets. Since  we
apply in our study the field-theoretical RG, Section \ref{II}
discusses the functional representation of the model of a random
anisotropy magnet for different types of distributions of the
local anisotropy axes as well as the stability conditions for the
effective Hamiltonians. Section \ref{III} presents the application
of the field-theoretical RG approach to study the critical
behaviour of the random anisotropy model (RAM). There, the RG
functions are obtained in the two-loop approximations within the
minimal subtraction scheme. The analysis of the {\it asymptotic}
and the {\it non-asymptotic} critical behaviour is performed using
the resummation techniques in space dimension $d=3$. Section
\ref{V} concludes our study and summarizes the results.

\section{A brief review of earlier experimental, numerical, and theoretical
studies\label{0}}

Magnetic properties of random anisotropy magnets are described by
the RAM \cite{Harris73}. This is a spin lattice model with each
spin subjected to a local anisotropy of random orientation. The
Hamiltonian reads:
\begin{equation}
{\mathcal H} =  - \sum_{{\bf R},{\bf R'}} J_{{\bf R},{\bf R'}}
\vec{S}_{\bf R} \vec{S}_{\bf R'} -D\sum_{{\bf R}} (\hat {x}_{\bf
R}\vec{S}_{\bf R})^{2}.
 \label{origham}
\end{equation}
Here,  $\vec{S}_{\bf R}$ is a  classical $m$--component vector
(``spin") on the  site ${\bf R}$ of a $d$--dimensional
(hyper)cubic lattice, $D>0$ is the strength of the anisotropy,
$\hat {x}_{\bf R}$ is a random unit vector pointing in the
direction of the local anisotropy axis. The interaction $J_{{\bf
R},{\bf R'}}$ is assumed to be ferromagnetic. For the short-range
case it can be written in form of the nearest-neighbours
interaction:
\begin{displaymath}
J_{{\bf R},{\bf R'}}=\left\{
\begin{array}{cc}
\bar{J}>0,&\quad |{\bf R}-{\bf R'}|=a;\\ 0& \quad
\mbox{otherwise},
\end{array}\right.
\end{displaymath}
where $a$ is the distance between the nearest neighbours. The
degree of disorder is controlled by the ratio $D/{\bar J}$.
Obviously, the random orientations are    present in
(\ref{origham}) only for $m>1$. In the case $m=1$ the random
anisotropy term becomes constant and leads only to a shift of the
Ising system free energy.

The RAM~(\ref{origham}) is relevant to describe a wide class of
disordered magnets. It was first introduced to describe magnetic
properties of amorphous alloys of rare-earth compounds with
aspherical electron distributions and transition metals
\cite{Harris73}. Today the majority of the amorphous alloys
including rare-earth elements is recognized to be random
anisotropy magnets. This class of amorphous materials includes
binary alloys
\cite{Coey81,Molnar82,Barbara85s,Dieny86,Lee93,Banjelloun99,Hassanain},
alloys of rare earths with two metallic components
\cite{Dieny85,Barbara85,Barbara87,Oshea86,Gruzalski79,Moral86,Sellmyer86,Moral92,Bewley95,Fuente96},
alloys of rare earth with a metal and a non-metallic component
\cite{Oshea83,Cornelison84,Tejada90,Driouch96,Slimani97} and
multicomponent alloys
\cite{Sellmyer85,Krishnan95,Krishnan96,Itri99,Bhandage97,Kato98}.
The structure of these compounds is characterized by an uniform
(isotropic) distribution of the direction of the random anisotropy
axes. Also crystalline compounds such as $RE_xY_{1-x}Al_2$
$(RE=Tb,\, Dy)$
\cite{Gehring90,Joven91,Moral94a,Moral94b,Moral95,Moral96a,Moral96b}
are   described by the RAM. However in this case, a distribution
of the random anisotropy axes orientations is confined to several
directions only. It turned out that random anisotropy plays a
distinguished role also in the magnetic behaviour of certain
uranium based amorphous alloys \cite{Homma98a,Homma98b,Itoh00}.
Although all materials mentioned above represent the case of spin
dimension $m=3$, certain magnets are described by the RAM with
$m=2$ \cite{Taylor96,Taylor01}.

Beside traditional magnets, random anisotropy is present in such
magnetic materials as molecular based magnets \cite{molmag},
nanocrystallity materials \cite{nano} and granular systems
\cite{granular}.

 The problem of the nature of a low-temperature phase of
random anisotropy magnets and peculiarities of the transition to
this phase was investigated extensively by theoretical approaches,
numerical simulations and experiments. Below we summarize what has
been obtained so far for the three dimensional systems. In a
separate subsection, we mention also some results for the RAM at
lower dimensions.

 \subsection{ Experiments} The earliest experimental data about amorphous alloys
containing rare earth and transition metals are collected in the
comprehensive review of Cochrane {\em et al.} \cite{Cochraine78}.
Results of later experimental investigations are reviewed in the
paper of Sellmyer and O'Shea \cite{rev}.  Most of the investigated
materials was recognized as systems with strong anisotropy. For
many of them in the  low-temperature phase an extremely large
magnetic susceptibility was observed which was consistent with
theoretically predicted peculiarities \cite{Aharony80}. However no
phase with a non-zero magnetization was obtained. The majority of
investigations of amorphous random anisotropy systems reports the
low-temperature phase to be  either a spin-glass phase or a so
called correlated spin glass phase, in terms of the
phenomenological theory \cite{ChudnovskySaslowSerota} often used
for the interpretation of data.

Most of the examples given above concern systems with an {\em
isotropic distribution} of the orientation of the random
anisotropy axes. However the crystalline compounds
$RE_xY_{1-x}Al_2$ $(RE=Tb,\, Dy)$
\cite{Gehring90,Joven91,Moral94a,Moral94b,Moral95,Moral96a,Moral96b}
are examples of  random anisotropy systems, where the distribution
of local anisotropy axes is {\em anisotropic}. The rich phase
diagram of these systems contrary to isotropic alloys presented
evidence of a  ferromagnetic phase for very weak anisotropy
\cite{Moral94a}.

Therefore the experiments indicate the absence of a ferromagnetic
phase for amorphous systems in general, with the possibility of
ferromagnetism for crystalline compounds with weak random
anisotropy. The effective critical behaviour of random anisotropy
magnets was not considered. The question about effective critical
exponents was touched only in Ref. \cite{Fahnle87}, where the
temperature region of asymptotic critical regime was discussed for
random anisotropy magnets.

\subsection{ Monte Carlo simulations} The majority of the computer
simulations exploiting the RAM concerned the case of strong
($D/{\bar J }>>1$) or more often  infinitely strong anisotropy
($D/{\bar J }\to\infty$). The earliest investigations report
inconclusive results: both  stability \cite{Chi77,Harris78} as
well as instability \cite{Chi79,Harris80} of the ferromagnetic
order with respect to the  spin-glass phase has been reported.
However, data of later investigations indicated the absence of
ferromagnetism. The restriction to the infinitely strong
anisotropy limit led to a lack of long-range order in the ground
state for $m=3$ \cite{Jayaprakash80}. In this limit, the critical
exponents at the transition to a low-temperature phase were found
to be similar to those of the three-dimensional short-range
interacting Ising spin glass \cite{Chakrabati87}. The results of
Monte Carlo calculations \cite{Fisch89,Fisch90} confirmed an
absence of  ferromagnetism.

On the other hand, Monte Carlo results for the RAM with $m=2$ are
consistent with the existence of a low-temperature phase of
extremely large susceptibility \cite{Fisch89}. This feature was
predicted theoretically
 for arbitrary $m$  and  weak anisotropy \cite{Aharony80}.
The possibility of the existence of such a phase  was indicated
also  in investigations of spin models with $p-$fold random fields
\cite{Fisch90}, which in the case $m=2$, $p=2$ correspond to the
RAM with $m=2$.

Monte Carlo simulations for $m=2$ show also the existence of a
non-magnetic low-temperatu\-re phase with power-law decay of a
pair correlation function predicted for the RAM \cite{Aharony80}.
The result \cite{Fisch91} for infinite anisotropy is consistent
with  this quasi-long-range ordered phase. On the other hand such
phase was not obtained for the model with $p-$fold fields  at
$\bar{J}=D$ \cite{Reed91}. Moreover, another Monte Carlo study of
the RAM with  $m=2$ and the same relation between $\bar{J}$ and
$D$ obtained critical exponents with values similar to the
$XY$-ferromagnetic transition, except that the heat capacity
critical exponent was found to be positive \cite{Rossler}. It was
shown that the long-range order is not destroyed in the
vortex-free model with $p-$fold fields \cite{Fisch00}.

The possibility of the existence of a quasi-long-ordered phase was
also obtained for $m=3$ in the case of weak anisotropy but
assuming $D/\bar{J}=\infty$  for a part $q$ of sites and
$D/\bar{J}=0$ for the rest of $1-q$ sites \cite{Fisch98}. Here,
the quasi-long-range ordered phase appeared as an intermediate
phase between the paramagnetic  and the ferromagnetic ordering.
However, for a non-zero anisotropy $D$ the low temperature phase
appears to be a quasi-long-range one \cite{Itakura03}.

To our knowledge, there exists only one the Monte Carlo study for
the random axis {\it anisotropic distribution}.
 This is a study of a cubic
model with random anisotropy, where the anisotropic axes are
oriented along the edges of a cube \cite{Fisch93}. For $m=2$ the
conventional XY second order phase transition to the ferromagnetic
phase was found for weak random anisotropy, whereas  a first-order
transition to a domain type ferromagnetic phase was found for
strong random anisotropy. For $m=3$ both transitions were found to
be of the first order.

Therefore, although there is a certain inconsistency of Monte
Carlo results, they mainly  confirm the experimentally observed
absence of ferromagnetism for random ani\-so\-tro\-py magnets with
{\it isotropic distribution} of random axes and bring about a
second order phase transition for the weak randomness with {\em
anisotropic distribution}. However, a low-temperature phase with
power-law decay of correlation functions as observed in Monte
Carlo simulations has not been found in the experiments so far.

\subsection{ Theoretical treatment} The first theoretical
investigations of the RAM were performed within  a mean field
theory.   Ferromagnetism was predicted  \cite{Harris73,mean} but
 the possibility of a spin-glass phase \cite{HarrisZobin77}
was not excluded. Within the mean field approach the
infinite-range interaction limit of the RAM was solved exactly
\cite{Derrida80}, indicating a second-order phase transition to
ferromagnetic order.

However, taking into account fluctuations lead the second order
phase transition disappeared. In the pioneer RG calculations
performed for RAM with {\em isotropic distribution} of $\hat
{x}_{\bf R}$ \cite{Aharony75} no stable accessible fixed point of
the RG transformation was found within the first order of
$\varepsilon=4-d$-expansion.
 Recently, this result was
corroborated by  two-loop \cite{Dudka01a,Dudka01c,Dudka02} and
even  five-loop \cite{new_Ital} calculations within the
field-theoretical RG refined by  resummation techniques.
Meanwhile, an effective free energy was derived for large $D$. It
was shown to have the same form as that of the random-bond Ising
spin glass \cite{Chen77}, demonstrating that the RAM can have a
spin glass phase.  Following the arguments of Imry and Ma
\cite{Imry75} formulated for the random-field model it was shown
that the $d=3$ random anisotropy magnet should break into magnetic
domains  of size $L\sim(\bar{J}/D)^2$ \cite{Alben78} for   weak
anisotropy and thus no ferromagnetism was expected. Several
different arguments were applied to the RAM \cite{Pelcovits78} in
order to demonstrate the absence of ferromagnetic order for space
dimensions $d<4$. Although among these arguments the one for the
limit $m\to\infty$ \cite{Pelcovits78} appeared to be erroneous
\cite{Pelcovits82}, the lack of a ferromagnetism for the RAM with
{\it isotropic distribution} of anisotropy axes for $d<4$ was
further supported by  a Mermin--Wagner type proof  using the
replica trick \cite{Emery75}. The same result was obtained  within
the Migdal--Polyakov RG technique \cite{Pelcovits79} avoiding the
application of the replica trick.

Investigating the equation of state of the RAM a zero
magnetization and an infinite magnetic susceptibility were
obtained in the low-temperature phase for any $m$
\cite{Aharony80}. Two-spin correlations in this phase possess a
power law decay.
 As mentioned above, the phase with
such a behaviour  is called quasi-long-range ordered. The estimate
of the susceptibility of low-temperature phase was corrected in
another paper \cite{Aharony83} and found to be finite with
$\chi\sim(D/\bar{J})^{-4}$ at $d=3$. A similar dependence of the
susceptibility was obtained by other approaches
\cite{Goldschmidt83,ChudnovskySaslowSerota}. The power law decay
of spin correlations in the low-temperature phase was corroborated
in particular for $m=2$ \cite{Villian84}. But the last result is
in disagreement with calculation results of Ref.
\cite{Dotsenko83}, where spin-glass phase was obtained. Applying
the functional renormalization group in $4-\varepsilon$ dimensions
\cite{Feldman99,Feldman00} to the RAM with $m=3$, quasi-long-range
order was found.

Investigations of the RAM in the spherical model limit
$m\to\infty$ were concentrated on the question about the
possibility of a spin glass phase. In this limit a ferromagnetic
order was obtained for $d>2$ and for $D$ less than some critical
value $D_c$, while for $D$ larger $D_c$ a spin glass phase was
obtained for arbitrary $d$ ($D_c=0$ for $d\le 2$)
\cite{Khorunzhy91}. For these investigations also  the
$1/m$-expansion
\cite{Goldschmidt83,Khurana84,Goldschmidt84,Jagannathan86} was
used. Applying the replica method a spin glass phase was found  to
exist below $d=4$ for arbitrary $D$ \cite{Goldschmidt83}. But this
spin glass solution later was shown to be unstable
\cite{Khurana84}. A stable spin glass phase for $d<4$ was obtained
avoiding the replica method \cite{Goldschmidt84}. However,
 the dynamics of spin glass order parameter
for the RAM showed an instability of the spin glass phase
\cite{Jagannathan86}. These results found their confirmation in
the mean field treatment of the $m\to\infty$ limit
\cite{Fisher91}, where the spin glass phase appeared only as a
feature of this limit and no spin glass phase was obtained for
finite $m$.

The phenomenological theory \cite{ChudnovskySaslowSerota} based on
the continuous version of the Hamiltonian (\ref{origham}) and
assuming correlations between randomly oriented an\-iso\-tro\-py
axes turned out  to be a more appropriate approach for the
interpretation of the field dependence of the experimentally
observed magnetization in the ordered phase. In this approach  the
spin correlation functions in different regimes
 of applied fields were analyzed \cite{ChudnovskySaslowSerota}. In particular, the
correlation length for small and zero fields was found to have the
form $\xi\sim R_a(\frac{\bar{J}}{R_a D})^2$, where $R_a$ is the
correlation range for random axes. Such a phase was called a
correlated spin-glass phase.

The infinitely strong anisotropy limit of the RAM  was
investigated
 with the help of high-temperature expansions. Results of
a Pad\'e-analysis \cite{Shender80} indicated typical spin glass
behaviour for space dimension $d=3$, while in   Ref.
\cite{Fisch90a} it was concluded that  the lower critical
dimension for  the RAM with $m=2$ is $d_L=3$. The high-temperature
series analysis of the Hamiltonian (\ref{origham}) on Cayley-trees
\cite{Harris87} predicted ferromagnetic order, occurring for the
number of nearest neighbours $\tilde{z}> m$ and a spin glass order
for $\tilde{z}< m$.

All theoretical approaches summarized above concerned the RAM with
{\em iso\-tro\-pic} random axes distribution. The {\em
anisotropic} case was first investigated  in a  RG study
\cite{Aharony75}  with a distribution  of anisotropic axes,
restricting directions of the axes  along the hypercube edges
({\em cubic distribution}). No accessible stable fixed point
corresponding to a second order phase transition point was found.
Investigating the RAM with mixed isotropic and cubic distributions
it was  found \cite{Fischer85} that the presence of a random cubic
anisotropy stabilizes the ferromagnetic phase. The possibility of
a second order phase transition into a ferromagnetic phase with
critical exponents of the diluted quenched Ising model for the RAM
with a cubic distribution was pointed out in Ref.
\cite{Mukamel82}. Such a behaviour was observed for a more general
model \cite{Korzhenevskii88} containing non-isotropic terms. It
included the RAM  with {\em cubic distribution} of random axes as
 a particular case. Subsequently, this result  was corroborated
 within a two-loop RG calculations with
resummation \cite{Dudka01c,Dudka02,Dudka01b} done directly for the
RAM and it was shown that the critical behaviour belongs to the
universality class of the site-diluted Ising model. This result
was further confirmed on the basis of a five-loop massive RG
calculations \cite{new_Ital}.

\subsection{RAM at low dimensions}

It is interesting to compare the above mentioned results with
those for low-dimensional random anisotropy systems. For $d{=}2$
both theoretical approaches \cite{Bray85} and numerical
simulations \cite{simul2} lead to a spin-glass character of the
low-temperature phase. An exception was a functional RG
investigation \cite{Feldman97} showing the instability of the
ferromagnetic order and a logarithmically slow decrease of the
correlation functions.

Results for the RAM in   $d=1$  dimensions do not give  such a
consistent picture.  Investigations of the $d=1$ infinity
anisotropy limit \cite{Thomas80} gave a spin-glass ground state.
However, a Monte Carlo investigation for finite $D$
\cite{Serota86} was unable to find a ground state with zero
magnetization. While another numerical study \cite{Dickman91}
demonstrated a decrease of the correlation length with $D$. Monte
Carlo simulations \cite{Reed93} indicated a zero temperature
magnetization.

\subsection{Some conclusions}

As one can see from the picture described above  there exists a
contradictory results concerning the transition of the random
anisotropy magnets into the low-temperature phase. The majority of
the results state that the $d=3$ RAM  has no low-temperature phase
with non-vanishing magnetization for the {\em isotropic
distribution} of the local anisotropy axes. Investigations of the
RAM with an {\em anisotropic distribution} of the anisotropy axes
bring about the possibility of  a ferromagnetic phase. Theory
predicts a power-low decay of correlation functions in the
low-temperature phase, which is confirmed by Monte Carlo
simulations, however no phase with such features was observed in
experiments. The question of {\em effective critical} behaviour of
random anisotropy magnets remains unclear.

Nowadays the field-theoretical RG approach serves as a reliable
method to get accurate results describing critical behaviour
\cite{rgbooks}. For the RAM,  only the papers
\cite{Aharony75,Dudka01a,Dudka01c,Dudka02,new_Ital,Mukamel82,Dudka01b}
were devoted to such analysis. They concentrated on the study of
the {\it asymptotic} critical behaviour of the RAM. Below we will
consider results of these studies in more details, revisiting the
$d=3$ RAM criticality within the minimal subtraction RG scheme.
Doing so, we will study  the {\em effective critical} behaviour of
the random anisotropy magnets. It is this non-asymptotic critical
behaviour that often is observed in experiments and Monte Carlo
simulations of critical systems near second order phase
transition. The analysis of the {\it asymptotic} and {\it
non-asymptotic} critical behaviour is carried out in the Section
\ref{III}, while in the next Section \ref{II} the effective
Hamiltonians for RAM with {\em isotropic} and {\em cubic
distributions} of local axes are discussed.

\section{Effective Hamiltonians for random anisotropy systems}\label{II}

The starting point for an RG analysis of the critical behaviour of
the spin model (\ref{origham}) is an effective Hamiltonian,
derived for a given random axes  $\{\hat x_{\bf R}\}$
distribution. The functional representation of the RAM and hence
the effective Hamiltonian  can be obtained using symmetry
considerations \cite{Aharony75,new_Ital}, however it can be
obtained also directly starting from the original spin Hamiltonian
(\ref{origham}). Along the line of derivation, it will become
clear, how the random axes distribution influences the symmetry of
the effective Hamiltonian and hence defines the universality
classes of our problem. The details of this procedure based on the
Stratonovich-Hubbard transformation are given in the Appendix.

The case we consider here corresponds to the non-equilibrium
disorder \cite{Brout59}: variables $\hat x_{\bf R}$ in
(\ref{origham}) are randomly distributed and fixed (quenched) in a
certain configuration. As derived in the Appendix, the
configuration-dependent partition function $Z(\{\hat x_{\bf R}\})$
for the case of weak anisotropy $D/\bar{J}<<1$ reads:

\begin{eqnarray}\label{zform}
Z(\{\hat x_{\bf R}\}) \sim \int(d\vec{\phi})e^{{\mathcal H}(\hat
x_{\bf R },\vec{\phi}_{\bf R})},
\end{eqnarray}
where the integral means functional integration in the space of
the $m$-com\-po\-nent fields $\vec{\phi}_{\bf R
}=({\phi}^{(1)}_{\bf R} \dots{\phi}^{(m)}_{\bf R} )$ and the
Hamiltonian ${\mathcal H}({\hat x_{\bf R},\vec{\phi}_{\bf R }})$
reads:
\begin{eqnarray}\label{ffrepresent}
{\mathcal H}({\hat x_{\bf R},\vec{\phi}_{\bf R }})&=&-\int d^d R
 \Big\{\frac{1}{2}\left[r_0|\vec{\phi}_{\bf R
}|^2+\left|\vec{\nabla}\vec{\phi}_{\bf R }\right|^2\right]{-}D_1
\left(\vec{\phi}_{\bf R}\hat x_{\bf
R}\right)^2{+}v_1|\vec{\phi}_{\bf
R}|^4+\nonumber\\&&z|\vec{\phi}_{\bf R}|^2\left(\vec{\phi}_{\bf
R}\hat x_{\bf R}\right)^2+\dots\Big\}.
\end{eqnarray}
Here, $r_0$ is proportional to the distance from the critical
temperature and $v_1$, $D_1$,  $z$ are positive constants
connected to the parameters of the spin Hamiltonian
(\ref{origham}) via relations explicitly given in the Appendix.

 This form of functional representation differs from the one
obtained by symmetry consideration \cite{Aharony75,new_Ital} by
the presence of the term with coupling $z$, however, as it will be
shown below, this term does not affect the critical behaviour of
the RAM since the symmetry of the resulting effective Hamiltonian
does not change when this term is omitted. In order to obtain the
free energy $F$ describing the physical system for the quenched
disorder \cite{Brout59} one has to average the logarithm of
configuration dependent partition function $Z(\{\hat x_{\bf R}\})$
over all possible random configurations $\{\hat x_{\bf R }\}$ of
the directions of the anisotropy axes:
\begin{equation}
-\beta F = \int\dots\int[d^m \hat x_{\bf R}] P(\{\hat x_{\bf
R}\})\ln Z(\{\hat x_{\bf R}\}).
\end{equation}
Here, $P(\{\hat x_{\bf R}\})$  is the probability of a realization
of a given configuration $\{\hat x_{\bf R}\}$. If one assumes that
there are no correlations between the directions of $\hat x_{\bf
R}$ on different sites, the probability distribution is factorised
into a product of distributions $p(\hat x_{\bf R})$ of $\hat
x_{\bf R}$ on each site $\bf R$ :
\begin{equation}
P(\{\hat x_{\bf R}\}) = \prod_{\bf R}p( \hat x_{\bf R}).
\end{equation}
To achieve the calculations of the logarithm of the partition
function it is convenient to use the replica trick \cite{Emery75}
introducing powers of $Z(\{\hat x_{\bf R}\})$ which are easier to
average:
\begin{equation}
\ln Z(\{\hat x_{\bf R}\}) = \left.\frac{\partial}{\partial
n}\left[Z(\{\hat x_{\bf R}\})\right]^n\right|_{n=0}.
\end{equation}

Now the model must be completed by choosing a certain distribution
$p(\hat x_{\bf R})$ for the random variables $\hat x_{\bf R}$. On
the one hand, this distribution should be simple enough from the
mathematical point of view, on the other hand, it must contain
certain physical constraints. Aharony \cite{Aharony75} considered
two types of distributions of $\hat x_{\bf R}$. The first is an
{\em isotropic} one, where the random vector $\hat x_{\bf R}$
points with equal probability in any direction in the
$m$-dimensional hyperspace:
\begin{equation}\label{isdist}
p(\hat x) \equiv\left(\int d^m \hat x\right)^{-1}=
\frac{\Gamma(m/2)}{2\pi^{m/2}}.
\end{equation}
Here $\Gamma(x)$ is Euler gamma-function, and the right-hand side
presents the volume of the $m$-dimensional hypersphere of unit
radius. This distribution mimics an amorphous system without any
preferred directions. The second distribution restricts the vector
$\hat x_{\bf R}$ to point with equal probability only along one of
the $2m$ directions of the axes $\hat k_i$ of a (hyper)cubic
lattice:
\begin{equation}\label{cubdist}
p(\hat x) = \frac{1}{2m}\sum_{i=1}^m\left[\delta^{(m)}(\hat x-\hat
k_i)+\delta^{(m)}(\hat x+\hat k_i)\right],
\end{equation}
where $\delta(y)$ are Kronecker's deltas. This distribution will
be called {\em cubic distribution} henceforth. The cubic
distribution corresponds to a situation when an amorphous magnet
still ``remembers" the initial cubic lattice structure in spite of
the random anisotropy or describes  crystalline compounds with
random but restricted anisotropy. It is an example of the cases of
more general anisotropic distributions.

Performing the average over the random variables $\{\hat x_{\bf
R}\}$ for the  {\em isotropic distribution} (\ref{isdist}) one
ends up with the  effective Hamiltonian \cite{Aharony75}:
\begin{eqnarray}\label{isham}
{\mathcal H}_{eff}& =& -\int d^d R \Bigg \{
\frac{1}{2}\left[\mu_0^2|\vec{\varphi}|^2+\left|\vec{\nabla}\vec{\varphi}\right|^2
\right]+u_0|\vec{\varphi}|^4+
\nonumber\\&&v_0\sum_{\alpha=1}^n|\vec{\phi}^{\alpha}|^4+
w_0\sum_{\alpha,\beta=1}^n\sum_{i,j}^m
\phi_i^{\alpha}\phi_j^{\alpha}\phi_i^{\beta}\phi_j^{\beta}.
\end{eqnarray}
It describes in the replica limit, $n\to0$, the critical
properties of  model (\ref{origham}) with distribution
(\ref{isdist}). Here $\mu_0=r_0-D_1/m$ is the bare
(non-renormalized) mass  and
 $u_0=\frac{D_1^2}{m^2(m+2)}>0$, $v_0=v_1+z/m>0$,
$w_0=-\frac{D_1^2}{m(m+2)}<0$ are the bare couplings.
$\phi_i^{\alpha}\equiv\phi_{i,{\bf R}}^{\alpha}$ are components of
the $n$-replicated $m$-dimensional field,
$|\vec{\varphi}|^2\equiv\sum_i
\sum_{\alpha}\left|\phi_i^{\alpha}\right|^2$. The ratio of the
couplings $w_0$ and $u_0$ equals $w_0/u_0=-m$ and determines a
region of physically allowed initial values in the $(u-v-w)$-space
of couplings.

For the {\em cubic distribution} (\ref{cubdist}) the average over
the random variables $\{\hat x_{\bf R}\}$ leads to the effective
Hamiltonian  of the following form \cite{Aharony75}:
\begin{eqnarray}\label{cubham}\!\!\!\! &{\mathcal H}_{eff}
 {=}& {-}\!\!\int\!\! d^d R \Bigg
\{
\!\frac{1}{2}\!\left[\mu_0^2|\vec{\varphi}|^2{+}\!\left|\vec{\nabla}\vec{\varphi}\right|^2
\right]\!\!{+}u_0|\vec{\varphi}|^4{+}v_0\!\!\sum_{\alpha=1}^n|\vec{\phi}^{\alpha}|^4\!{+}\nonumber\\
&& w_0\!\sum_{i=1}^m\!\sum_{\alpha,\beta{=}1}^n
\left(\phi_i^{\alpha}\right)^2\left(\phi_i^{\beta}\right)^2{+}
y_0\sum_{i=1}^m\sum_{\alpha=1}^n\left(\phi_i^{\alpha}\right)^4\!.
\end{eqnarray}
Again, one has to perform the replica limit $n\to0$ to get the
physical quantities. Here, the bare mass and couplings are
determined as: $\mu_0=r_0-D_1/m$, $u_0=\frac{D_1^2}{2m^2}>0$,
$v_0=v_1+z/m>0$, $w_0=-\frac{D_1^2}{2m}<0$.  The last term in
(\ref{cubham})  has cubic symmetry.
 It does not result from the steps
leading to ${\mathcal H}_{eff}$ but has to be included in
(\ref{cubham}) since it is generated by further application of the
RG transformation. Therefore $y_0$ can be of either sign. The
symmetries of $w_0$ terms differ in (\ref{isham}) and
(\ref{cubham}). Furthermore, although values of $u_0$ and $w_0$
differ for Hamiltonians (\ref{isham}) and (\ref{cubham}), their
ratio again equals $-m$.

It should be noted that for $m=1$ the effective Hamiltonians
(\ref{isham}) and (\ref{cubham}) reduce to the traditional
effective Hamiltonian of the Ising magnet with one coupling,
demonstrating the degeneration of the random anisotropy term to a
constant  for the Ising spins, as  mentioned in the introduction.
Indeed, in this case the $u_0$- and $w_0$-terms turn out to have
the same symmetry (as well as the $v_0$- and $y_0$-terms in
(\ref{cubham})). Moreover, due to the fixed ratio of these
couplings they are equal in their absolute values but have
opposite signs and therefore cancel one another. This leads
finally to the usual $\phi^4$-Hamiltonian.

Other distributions leading to similar effective Hamiltonians as
described abo\-ve are discussed in the Ref. \cite{new_Ital}.   In
particular, let the distribution $p(\hat x)$, have moments:
 $M_{i_1 i_2 \dots i_k}\equiv\int d^n x\, p(\hat
x)\,x_{i_1}x_{i_2}\dots x_{i_k}$
 satisfying the conditions:
\begin{eqnarray}\label{d_cond}
M_{ab}=\frac{1}{m}\delta_{ab},\qquad
M_{abcd}=A(\delta_{ab}\delta_{cd}+\delta_{ac}\delta_{bd}
+\delta_{ad}\delta_{bc})+B\delta_{ab} \delta_{ac}
\delta_{ad},\end{eqnarray}

with Cauchy inequalities $A(m+2)+B\ge {1}/{m}$ and
$3A+B\ge{1}/{m^2}$ for the parameters $A$ and $B$. Then the
effective Hamiltonian of the RAM with random anisotropy axis
distribution with moments (11) turns out to have the form:
\begin{eqnarray}\label{fham}
{\mathcal H}_{eff}& =& -\int d^d R \Bigg \{
\frac{1}{2}\left[(r_0-D_1/m)|\vec{\varphi}|^2+\left|\vec{\nabla}\vec{\varphi}\right|^2
\right]+v_0\!\!\sum_{\alpha=1}^n|\vec{\phi}^{\alpha}|^4+\nonumber\\&&
\frac{D_1^2}{2m^2}(1-m^2A)|\vec{\varphi}|^4-AD_1^2\sum_{\alpha,\beta=1}^n\sum_{i,j}^m
\phi_i^{\alpha}\phi_j^{\alpha}\phi_i^{\beta}\phi_j^{\beta}-\nonumber\\&&
\frac{BD_1^2}{2}\!\sum_{i=1}^m\!\sum_{\alpha,\beta{=}1}^n
\left(\phi_i^{\alpha}\right)^2\left(\phi_i^{\beta}\right)^2.
\end{eqnarray}
However,   a term
$\sum_{i=1}^m\sum_{\alpha=1}^n\left(\phi_i^{\alpha}\right)^4$
 has to be added here for the same reason, as it was
included in (\ref{cubham}). Thus the effective Hamiltonian
corresponding to the distribution of $p(\hat x)$ with conditions
(\ref{d_cond}) reads:
\begin{eqnarray}\label{fulham}\!\!\!\! &\!\!\!{\mathcal H}_{eff}
 {=}&\! {-}\!\!\int\!\! d^d R \Bigg
\{
\!\frac{1}{2}\!\left[\mu_0^2|\vec{\varphi}|^2{+}\!\left|\vec{\nabla}\vec{\varphi}\right|^2
\right]\!\!{+}u_0|\vec{\varphi}|^4{+}v_0\!\!\sum_{\alpha=1}^n|\vec{\phi}^{\alpha}|^4\!{+}\nonumber\\
&& w_0\!\sum_{i=1}^m\!\sum_{\alpha,\beta{=}1}^n
\left(\phi_i^{\alpha}\right)^2\left(\phi_i^{\beta}\right)^2{+}
y_0\sum_{i=1}^m\sum_{\alpha=1}^n\left(\phi_i^{\alpha}\right)^4\!+
{z_0}\!\!\!\sum_{\alpha,\beta=1}^n\sum_{i,j}^m
\phi_i^{\alpha}\phi_j^{\alpha}\phi_i^{\beta}\phi_j^{\beta},
\end{eqnarray}
Such a Hamiltonian contains symmetry terms of both effective
Hamiltonians (\ref{isham}) and (\ref{cubham}). It was originally
introduced in Ref. \cite{Mukamel82} to describe magnetic systems
with single-ion anisotropy and non-magnetic impurities. The
Hamiltonian (\ref{fulham}) was studied in the first order in
$\varepsilon$ and no accessible and stable fixed point was found
\cite{Mukamel82}. One should note that the $y_0$- term  in the
(\ref{fham}) as well as in (\ref{cubham}) appears in a natural
way, if one considers the original
 spin model (\ref{origham}) to contain a cubic symmetry  term $\sum_{\bf R}\sum_{i=1}^m
S^4_{i\,\bf R}$.

For distributions 
 with $A=0$ in (\ref{d_cond}),  the effective Hamiltonian (\ref{fulham})
reduces simply to (\ref{cubham}), since  $\tilde{w_0}=0$. As it
was noted in Ref. \cite{new_Ital} this feature is characteristic
for all distributions, which orient the anisotropy axis only along
the axes of an $m$-dimensional hypercube, including the cases when
the anisotropy axis points in positive or in negative directions
only.

For the further analysis it is instructive to establish the
regions in the space of the coupling constants where the free
energy does not diverge in the ordered phase or in
 other words where the  Hamiltonians (\ref{isham}) and
(\ref{cubham}) are stable. The regions vary  for different ordered
phases, described by certain nonvanishing values of the order
parameter $M$.  If one neglects field fluctuations and if the
symmetry of the ordered phase is defined by (i)
$\phi_i^{\alpha}=M$ then the regions of stability  for the
effective Hamiltonian (\ref{isham}) within the space of couplings
are defined by the inequality \cite{Mukamel82}:
\begin{eqnarray} &&\label{c1}v_0+nu_0+nw_0>0.
\end{eqnarray}
For the case (ii) $\phi_i^{\alpha}=M\delta_{\alpha 1}\delta_{i 1}$
the following condition has to be fulfilled \cite{Mukamel82}:
\begin{eqnarray} &&\label{c2} v_0+u_0+w_0>0.
\end{eqnarray}
 The  symmetries (iii) $\phi_i^{\alpha}=M\delta_{i 1}$ and
(iv) $\phi_i^{\alpha}=M\delta_{\alpha 1}$ lead to the same
conditions (\ref{c1}) and (\ref{c2}) respectively.
  In a similar
way  four stability conditions are found  \cite{new_Ital} for the
effective Hamiltonian  (\ref{cubham}):
\begin{eqnarray}
&{\rm (i)}&\qquad \label{cc1}mnu_0+mv_0+nw_0+y_0>0,\\ &{\rm
(ii)}&\qquad u_0+v_0+w_0+y_0>0,\\ &\label{cc2}{\rm (iii)} &\qquad
nu_0+v_0+nw_0+y_0>0,
\\
 & {\rm (iv)}&\qquad mu_0+mv_0+w_0+y_0>0.
 \end{eqnarray}
As it was argued in  Ref. \cite{Mukamel82}, the legitimate
stability conditions in the replica limit are those for which
isotropy holds in the replica space. Then for Hamiltonian
(\ref{isham}) only the inequality (\ref{c1}) is a relevant
stability condition, while for Hamiltonian (\ref{cubham}) the
appropriate conditions are inequalities (\ref{cc1}) and
(\ref{cc2}). In the replica limit they reduce to conditions:
\begin{eqnarray}\label{st_con}
v_0>0
 \end{eqnarray}
and
\begin{eqnarray}\label{st_cond}
&&mv_0+y_0>0, \qquad v_0+y_0>0. \end{eqnarray}

Together with  the condition $w_0/u_0=-m$ the above inequalities
(\ref{st_con}) and (\ref{st_cond}) determine all possible initial
values of the effective Hamiltonians that can be used in the RG
analysis, compared to some expiremental situation or computer
simulations.

\section{Critical behaviour of the RAM as explained by the renormalization group}\label{III}

In this section  the field-theoretical RG approach is applied to
analyse critical properties of the RAM. Our previous
field-theoretical RG studies exploited the massive renormalization
scheme \cite{Dudka01a,Dudka01c,Dudka02,Dudka01b} and were
restricted to investigation of asymptotic critical properties.
Already in the two-loop approximation
\cite{Dudka01a,Dudka01c,Dudka02,Dudka01b} the critical behaviour
was explained, then it was considered again in a five-loop study
\cite{new_Ital}
 confirming the former results. Here, we use the minimal subtraction RG scheme within
the two-loop approximation to study the effective critical
behaviour of the $d=3$ RAM \cite{Dudkathesis}. Since all essential
features of asymptotic criticality become evident in this
approximation, we consider it to be  sufficient  for analysing the
effective critical behaviour.

\subsection{Renormalization}\label{III.0}

The  field-theoretical RG approach \cite{rgbooks} is based on the
renormalization of one-particle irreducible (1IP) vertex functions
defined as
\begin{eqnarray}\label{gumfunc}
&\delta (\sum_i^{L} p_i{+}\sum_j^{N}
k_j)\Gamma_0^{(L,N)}(\{p\},\{k\},\mu_0^2,\{u_{i,0}\},\Lambda_0){=}\int^{\Lambda_0}_0\!
e^{i(\sum_ip_ir_i{+}\sum_jk_jR_j)}&\nonumber\\
&<\!\phi^2(r_1)\dots\phi^2(r_L)\phi(R_1)\dots\phi(R_N)\!>_{\rm 1PI
}^{{\mathcal H}_{eff}}\!dr_1\dots dr_LdR_1\dots dR_N,&
\end{eqnarray}

where $\{u_{i,0}=u_0,v_0,w_0,y_0,...\}$ is a set of bare
 couplings, $\mu_0$ is a bare mass, $\{p\}$ and
$\{k\}$ are sets of the external momenta and $\Lambda_0$ is the
cut-off parameter. The angular brackets in (\ref{gumfunc}) denote
the statistical average over the Gibbs distribution with the
effective Hamiltonian (\ref{isham}) or (\ref{cubham}).

The vertex functions appear to be divergent in the infrared limit
$\Lambda_0\to\infty$. To remove the divergencies, a controlled
rearrangement of the perturbative series for the vertex functions
is performed. The finiteness of renormalized vertex functions
$\Gamma^{(L,N)}$ is ensured by imposing certain normalizing
conditions. This leads to different  renormalization schemes. In
the following, the dimensional regularization with minimal
subtraction is used \cite{tHooft72}.

Normalizing conditions for the minimal subtraction scheme
\cite{tHooft72} are imposed at zero mass and have the following
form:
\begin{eqnarray}\label{mincond}
\left.\Gamma^{(0,2)}(k,-k;\tilde\mu;\{u_i\})\right|_{k=0}&=&0,\nonumber\\
\frac{d}{d\,k^2}\left.\Gamma^{(0,2)}(k,-k;\tilde\mu;\{u_i\})\right|_{k^2=\tilde\mu^2}&
=&1,\nonumber\\ \left.
\Gamma^{(0,4)}_{u_i}(\{k\};\tilde\mu;\{u_i\})\right|_
{k_{i}{\bf\cdot} k_{j}=\frac{\tilde\mu^2}{3}(4\delta_{ij}-1)} &=&
\tilde\mu^{\varepsilon}u_i,\\
\left.\Gamma^{(1,2)}(p;k,-k;\tilde\mu;\{u_i\})\right|_{\scriptsize\begin{array}{c}
p^2=k^2=\tilde\mu^2\\ pk=-1/3\tilde\mu^2
\end{array}}&=&1.\nonumber
\end{eqnarray}
Here, $u_i=\tilde{\mu}^{\varepsilon}\frac{Z_{\phi}^2}{Z_{u_i}}u_0$
are renormalized couplings and  $\tilde\mu$ is the external
momentum scale parameter, $\varepsilon=4-d$.
 Expansions of the renormalizing factors of the
field $\phi$, $Z_{\phi}$, the couplings $\{u_{i}=u,v,w,y,...\}$,
$Z_{u_i}$, and for the operator $\phi^2$, ${\bar Z_{\phi^2}}$,
ensuring finiteness of two-point vertex function with one $\phi^2$
insertion,  can be obtained from expressions (\ref{mincond}). The
 functions $\Gamma^{(0,4)}_{u_i}$ are coefficients of terms of different symmetries
in the expression for the four-point vertex function
$\Gamma^{(0,4)}{}^{ijkl}_{\alpha\beta\gamma\tau}$,  containing the
full tensorial structure.
 The last function for the {\em isotropic distribution} reads:
\begin{equation}
\Gamma^{(0,4)}{}^{ijkl}_{\alpha\beta\gamma\tau}{=}
\Gamma^{(0,4)}_{u}{S}^{ijkl}_{\alpha\beta\gamma\tau} {+}
\Gamma^{(0,4)}_{v}{S}_{ijkl}F_{\alpha\beta\gamma\tau} {+}
\Gamma^{(0,4)}_{w}{A}^{ijkl}_{\alpha\beta\gamma\tau},
\end{equation}
while for the {\em cubic distribution} it is given by
\begin{eqnarray}
\Gamma^{(0,4)}{}^{ijkl}_{\alpha\beta\gamma\tau}&{=}&
\Gamma^{(0,4)}_{u}{S}^{ijkl}_{\alpha\beta\gamma\tau} {+}
\Gamma^{(0,4)}_{v}{S}_{ijkl}F_{\alpha\beta\gamma\tau}
{+}\nonumber\\
&&\Gamma^{(0,4)}_{w}{F}_{ijkl}S_{\alpha\beta\gamma\tau}{+}
\Gamma^{(0,4)}_{y}{F}_{ijkl}F_{\alpha\beta\gamma\tau},
\end{eqnarray}
where
\begin{eqnarray}
{F}_{ijkl}&{=}&\delta_{ij}\delta_{ik}\delta_{il},\nonumber\\
{S}_{ijkl}&{=}&\frac{1}{3}\left(\delta_{ij}\delta_{kl}+
\delta_{ik}\delta_{jl}+\delta_{il}\delta_{jk}\right),\nonumber\\
{S}_{ijkl}^{\alpha\beta\gamma\tau}&{=}&\frac{1}{3}\left(\delta_{ij}\delta_{kl}
\delta_{\alpha\beta}\delta_{\gamma\tau}{+}
\delta_{ik}\delta_{jl}\delta_{\alpha\gamma}\delta_{\beta\tau}
{+}\delta_{il}\delta_{jk}\delta_{\alpha\tau}\delta_{\beta\gamma}
\right),\nonumber\\  {A}_{ijkl}^{\alpha\beta\gamma\tau}&{=}&
{S}_{ijkl}{S}_{\alpha\beta\gamma\tau}-{S}_{ijkl}^{\alpha\beta\gamma\tau},
\end{eqnarray}
and $\delta_{ab}$ is the Kronecker's delta.

The renormalized vertex functions satisfy the following
homogeneous RG equation \cite{rgbooks}:
\begin{eqnarray}\label{rge}
&\Big\{ \tilde\mu\frac{\partial}{\partial \tilde\mu}{+}
\sum_i\beta_{u_i}(\{u_i\})\frac{\partial}{\partial u_i}{-}
\frac{N}{2}\gamma_{\phi}{+}&\nonumber\\ &
L\left({\bar\gamma_{\phi^2}}{+}\gamma_{\phi}\right) \Big\}
\Gamma^{(L,N)}(\{p\};\{k\};\tilde\mu;\{u_i\})=0,&
\end{eqnarray}
(except for the case $L=2$ and $N=0$) with the coefficients:
\begin{eqnarray}\label{beta}
\beta_{u_i}&{=}&\frac{\partial\,u_i}{\partial \ln \tilde{\mu} },\\
\label{gam1} {\gamma}_{\phi}&{=}&\frac{\partial\ln
Z_{\phi}}{\partial \ln \tilde{\mu} },\\ \label{gam2}
 \bar{\gamma}_{\phi^2}&{=}&-\frac{\partial\ln \bar{Z}_{\phi^2}}{\partial\ln\tilde{\mu} },
\end{eqnarray}
determining the approach of a system to criticality. The case
$L=2$ and $N=0$ fulfils an inhomogeneous equation and renormalizes
additively. It is important for calculating the specific heat.
Because of the scaling laws two exponent functions
${\gamma}_{\phi}$, $\bar{\gamma}_{\phi^2}$ are enough to get all
exponents. This also holds for the effective exponents above $T_c$
in the approximation we use them here.

The solution of the RG equation (\ref{rge}) via the method of
characteristics introduces the flow equations:
\begin{equation}\label{floweq}
\ell\frac{d u_i}{d\ell}=\beta_{u_i}(\{u_i\}).
\end{equation}
The flow parameter $\ell$ is related via a matching condition to
the distance from the critical point $\tau$. In the limit $\ell
\to 0$ the scale-dependent values of the couplings $u_i(\ell)$
will approach their stable fixed point (FP) values $\{u^*_i\}$, if
such a FP exists and if it is attainable from the initial values
for (\ref{floweq}) (initial ``background" values of couplings).

The FPs  of the system of equations (\ref{beta}) are defined as
the set of  real zeros of the $\beta$-functions:
\begin{equation}\label{FP}
\beta_{u_i}(\{u^*_i\}){=}0.
\end{equation}
A FP is stable if all eigenvalues $\omega_i$ of the stability
matrix $B_{i,j}=\frac{\partial\beta_{u_i}}{\partial {u_j}}$
calculated at this FP have  positive real parts. A stable FP
corresponds to the critical point of the system only if this FP is
reachable from physical meaningful initial conditions. The
critical exponents are then obtained from the $\gamma$-functions
evaluated at this stable  FP. The corresponding expressions for
the exponent $\nu$ of the correlation length  and the exponent
$\eta$ of the pair correlation function at $T_c$ are expressed by
the FP values of the $\gamma$-functions (\ref{gam1}),
(\ref{gam2}):
\begin{eqnarray}\label{nu}
\nu&{=}&\left [
2-\bar{\gamma}_{\phi^2}(\{u^*_i\})-{\gamma}_{\phi}(\{u^*_i\})\right]^{-1},\\
 \eta&=&{\gamma}_{\phi}(\{u^*_i\})\label{eta}.
\end{eqnarray}
The solution of the RG equation (\ref{rge}) together with the
solution for vertex function $\Gamma^{(2,0)}$ proof scaling.
Therefore, all other critical exponents may be obtained from the
familiar scaling laws. For example, the susceptibility exponent
$\gamma$ is given by:
\begin{equation}\label{slaw}
\gamma{=}(2-\eta)\nu,\end{equation}
 and thus expressed by the $\gamma$-functions (\ref{gam1}), (\ref{gam2}) as:
\begin{equation}\label{gamma}
\gamma^{-1}{=}1-\frac{\bar{\gamma}_{\phi^2}(\{u^*_i\})}{2-{\gamma}_{\phi}(\{u^*_i\})}.
\end{equation}
Other exponents are found in a similar way.

\subsection{Asymptotic critical behaviour \label{III.II}}
First, we analyse the asymptotic critical behaviour of the
effective Hamiltonians (\ref{isham}) and (\ref{cubham}) within the
renormalization scheme described above. For this purpose we write
down the explicit expressions for the two-loop RG functions. The
resulting   FPs  obtained by solving the system of the equations
for the FP (\ref{FP}) are compared with the results of the massive
scheme. This is the basic step before studying  the effective
critical behaviour in the next subsection \ref{III.III}.

\subsubsection{Isotropic distribution}\label{III.1}

Imposing the renormalization conditions of the minimal subtraction
scheme (\ref{mincond}) to the vertex functions of the theory with
the effective Hamiltonian (\ref{isham}) we found the $\beta$- and
$\gamma$-functions within the two-loop approximation (that is
within the second and third order in couplings for the $\beta$-
and $\gamma$-functions correspondingly):
\begin{eqnarray}\label{betais1}
\beta_u&=&{-}u\varepsilon{+}\frac{mn+8}{6}{u}^{2}{+}\frac{m+2}{3}uv{+}\frac{m{+}n{+}1}{3}uw
{+}\frac{1}{3}wv{ +}\frac{1}{2}{w}^{2}{-}\nonumber\\&& \frac
{3mn{+}14 }{12}{u}^{3}{-} {\frac {11\left (m+2\right
)}{18}}{u}^{2} v{-}{\frac {5\left (m{+}2\right )}{36}}
{v}^{2}u{-}\nonumber\\&&{\frac {\left (5mn{+}5m{+}5{n}+87\right
)}{72}}{w}^{2}u {-}{\frac {11\left (m{+}n{+}1\right )}{18}}
\!{u}^{2}w{-} \nonumber\\&&{\frac {5m{+}22}{18}}uwv{-}\frac
{m{+}8}{18}{w}^{2}v{-}\frac{m+n+3}{12}
{w}^{3}{-}\frac{1}{9}{v}^{2}w,
\end{eqnarray}

\begin{eqnarray}\label{betais2}
\beta_v&=&{-}\varepsilon v{+}\frac{m+8}{6}{v}^{2}{+}2vu{+}\frac
{m+5}{3}vw {-} \nonumber\\ &&\frac{3m{+}14}{12}{v}^{3}{-}{ \frac
{11m{+}58}{18}}{v}^{2}u{-}{\frac {17m{+}5n{+}65}{ 18}}wvu{-}
\nonumber\\ &&{\frac{5mn{+}82}{36}}v{u}^{2}{-}{\frac
{13m+56}{18}}w{v}^{2}{-}\frac{mn+33m+9n+131}{72}{w}^{2}v,
\end{eqnarray}

\begin{eqnarray}\label{betais3}
\beta_w&=&{-}\varepsilon
w{+}\frac{m{+}n{+}4}{6}{w}^{2}{+}2uw{+}\frac{2}{3}wv -\nonumber\\
&&{\frac {mn{+}3m{+}3n{+}17}{ 24}}{w}^{3}{-}{\frac
{11m{+}11n{+}29}{18}} {w}^{2}u{-}{\frac {5mn+82}{36}}w{u}^{2}
{-}\nonumber\\&&{\frac {m{+}10}{36}}w{v}^{2} {-}{\frac {2\left
(m{+}5\right )}{9}}{w}^{2}v  {-}{\frac {5m{+}34}{18}}wvu,
\end{eqnarray}

\begin{eqnarray}\label{gammais1}
\gamma_\phi&{=}&0+\nonumber\\&&{\frac{m+2}{36}}wv+\frac{m+n+1}{36}uw+
\frac{m+2}{36}vu{+}\nonumber\\&&{\frac
{mn{+}m{+}n{+}3}{144}}{w}^{2}{+}\frac  {m{+}2} {72}{v}^{2} +{\frac
{mn{+}2}{72}}{u}^{2}
\end{eqnarray}

\begin{eqnarray}\label{gammais2}
\bar{\gamma}_\phi^2&=&\frac {mn+2}{6}u+\frac
{m+2}{6}{v}+\frac{m+n+1}{6}w -\nonumber\\
&&\frac{mn+2}{12}{u}^{2}-\frac{m{+}2}{6}vu{-}\frac{m{+}2}{12}{v}^{2}{-}\frac{m+2}{6}vw
{-}\nonumber\\&&\frac{m{+}n{+}1}{6}uw{-}\frac{mn+m+n+3}{24}w^{2}.
\end{eqnarray}

Here, $\varepsilon=4-d$ and $u,\,v,\,w$ are renormalized
couplings. We recall that $m$ is the spin component number and $n$
is the number of replicas. The first line of each formula gives
the one loop result in the corresponding RG function. We present
the RG functions for any $m$ and $n$, although in the further
analysis of the RAM one considers the replica limit $n=0$.
However, the effective Hamiltonian (\ref{isham}) for non-zero $n$
may have also other physical interpretations.

An analysis of the RG functions (\ref{betais1})-(\ref{gammais2})
can be performed in two complementary ways. First, one may apply
the $\varepsilon$-expansion
 \cite{Wilson72}. Second, one may   fix $\varepsilon$
(i.e. space dimension $d$) in (\ref{betais1})-(\ref{gammais2})
 and solve equations (\ref{betais1})-(\ref{betais3}) for the FP numerically \cite{Schloms87}. In
 first order of the $\varepsilon$-expansion,  no {\em accessible}
stable FP was obtained \cite{Aharony75}. In  second order,
calculations performed within the massive scheme
\cite{Dudka01a,Dudka01c,Dudka02} did not change this picture
qualitatively. Application of the $\varepsilon$-expansion within
the minimal subtraction scheme leads to FPs coordinates which,
being  dependent on the renormalizing conditions \cite{rgbooks},
differ from those obtained within the massive scheme
\cite{Dudka01a}. The expansions of universal quantities, e. g.
critical exponents, coincide in both schemes. Leading to a
quantitative information, these expansions for the disordered
models \cite{Pelissetto02,Folk03} in general do not give  reliable
numerical values for the critical exponents. To give an example,
in the case of a random Ising model the expansion turns out to be
in  $\sqrt{\varepsilon}$, rather then ${\varepsilon}$
\cite{Grinstein76,sqrt} and does not lead to reliable estimates at
$\varepsilon=1$ ($d=3$) \cite{Folk00}. Therefore we do not display
here the results of the expansion in dimension $\varepsilon$ but
we discuss in more details the results  within the fixed dimension
$d$ approach.

The weak-coupling expansions obtained within perturbation theory
for the RG functions are asymptotic at best \cite{rgbooks}. In
order to obtain reliable results one has to apply  resummation
procedures. Here, the Pad\'e-Borel resummation technique
\cite{Baker78} is used. It consists in the following steps. For a
given initial polynomial in several variables (in this case in
three) for any finite series of $\beta=\beta_{u_i}$
\begin{equation}
\beta(u,v,w)=\!\sum_{1\le i+j+k\le 3}\!a_{i,j,k}\,u^iv^jw^k
\end{equation}
one introduces a ``resolvent" polynomial  in one auxiliary
variable $\lambda$ by \cite{Watson74}:
\begin{equation}
\label{r1} F(u,v,w;\lambda)=\!\sum_{1\le i+j+k\le
3}\!a_{i,j,k}\,u^iv^jw^k \lambda^{i+j+k-1},
\end{equation}
with the obvious relation $F(u,v,w;\lambda=1)=\beta(u,v,w)$. Then,
the Borel image of (\ref{r1}) is defined as:
\begin{equation}
\label{r2} F^B(u,v,w;\lambda)=
 \sum_{1\le i+j+k\le 3}\!
\frac{a_{i,j,k}{u}^i{v}^j{w}^k\lambda^{i+j+k-1}}{(i+j+k-1)!}.
\end{equation}
This truncated series (\ref{r2})  is approximated by the
Pad\'e-approximant $[1/1](\lambda)$ \cite{Baker81}.  Then, the
resummed $\beta$-function is obtained from the formula:
\begin{equation}
\label{r3} \beta^{res}(u,v,w)=\int_0^{\infty}dt\exp(-t)[1/1](t).
\end{equation}

Applying the above procedure for the analysis of the RG functions
(\ref{betais1})-(\ref{betais3}) at fixed space dimension $d=3$ and
solving the FP equations (\ref{FP}) we get nine FPs. Table
\ref{tabmin} lists  the FPs for $m=2,\,3,\,4$ in the region of
interest: $u^*>0$, $v^*>0$, $w^*<0$. The numeration of FPs
corresponds
 to that of Refs. \cite{Aharony75,Dudka01a,Dudka01c,Dudka02}.  FPs
 V and VII are not listed in the Table \ref{tabmin} since their coordinates are not in
 the region of interest.
 Performing the resummation procedure (42)-(45) sometimes one is lead
to Pad\'e-approximants containing a pole in the region of
integration in (45). Then the principal value of the integral is
taken. Coordinates of the fixed points obtained in this way are
shown in the Table 1 by small numbers. Some of the FPs in Table
\ref{tabmin} correspond to  field-theoretical models, relevant for
specific physical systems. In particular,  FPs II-IV recover
results for the isotropic $m$-vector, polymer ($m=0$), and the
diluted $m$-vector model correspondingly. All FPs in the
Table~\ref{tabmin} were already found within the
$\varepsilon$-expansion \cite{Aharony75}. Their presence was
further confirmed by the massive scheme of the field-theoretical
RG \cite{Dudka01a,Dudka01c,Dudka02}.
  Among these FPs only FP III is
the stable one, however this FP is not accessible from the initial
conditions for the coupling constants.
\begin{table}
\begin{center}
 {\begin{tabular}{|c|c|c|c|c|c|c|} \hline
FP&$m$&$u^*$&$v^*$&$w^*$&$\nu$&$\gamma$\\
 \hline I &$\forall m$& 0& 0& 0&1/2 &1 \\
\hline &2& 0 & 1.1415 & 0 & 0.684 &1.344 \\ II&3&0 & 1.0016 & 0 &
0.720 &1.411\\ &4& 0 & 0.8878 & 0 & 0.751 & 1.470\\ \hline
III&$\forall m$& 1.5281 & 0 & 0 & 0.599 & 1.185 \\ \hline
&2&0.1429 & 0.9923 & 0 & 0.665 & 1.307\\ IV&3&0.3411 & 0.6965 & 0
& 0.660 & 1.298\\ &4& 0.4546 & 0.5323 & 0 & 0.656 & 1.291\\ \hline
VI&2&1.6637 & 0 & -1.8212 & 0.452 & 0.911\\ &3&$^{1.1351}$ & 0 &
$^{-0.9073}$ & 0.470 & 0.944\\ \hline &2& 1.0126 & 0.9058 &
-0.6522 & 0.631 & 1.243\\ VIII&3&1.0728 & 0.7310 & -0.5483 & 0.626
&1.235\\ &4&1.1125 & 0.6241 & -0.4823 & 0.624 & 1.230\\ \hline
\end{tabular}}
\caption{\label{tabmin}  Fixed points and critical exponents $\nu$
and $\gamma$ as obtained on the base of resummed two-loop $\beta$-
and $\gamma$-functions for the RAM with {\em isotropic
distribution} of random anisotropic axes.  The only FP III is the
stable one. }
\end{center}
\end{table}

Table  \ref{tabmin} also presents the correlation length critical
exponent $\nu$ and the magnetic susceptibility critical exponent
$\gamma$ calculated for at the FPs. They are found from
expressions
$\nu^{-1}=2-\gamma_{\phi}(\{u^*_i\})-\bar{\gamma}_{\phi^2}(\{u^*_i\})$
and $\gamma^{-1}=1-\bar{\gamma}_{\phi^2}(\{u^*_i\})
/(2-\gamma_{\phi}(\{u^*_i\}))$, which are resummed in the same way
as the $\beta$-functions \cite{note}.

It should be noted that another stable FP which does not have an
analogue among the FPs presented for the massive scheme
\cite{Dudka01a,Dudka01c,Dudka02} was found in the region of
interest. We do not list the FP in the Table \ref{tabmin} since we
consider it uphysical for  the reasons explained below. This new
FP is not in correspondence with any FP found within
$\varepsilon$-expansion \cite{Aharony75}. The presence of this FP
can be an artifact of the resummation procedure caused by the
nonlinearity of the system of the FP equations. Thus, the
resummation can lead to the appearance of unphysical FPs. A
possible check if a FP makes  physical sense can be done by
following the dependence of its coordinates on  space dimension
$d$. If the non-trivial FP at $d=3$ can be traced back to the
non-trivial solution at $d=4$ (where the theory is known to
contain only the Gaussian FP \cite{rgbooks}) this may serve as
strong evidence of the unphysical nature of such a FP. However the
dimensional dependence of the  coordinates of the FP in question
could be calculated only up to $d=3.77$, then no real solution can
be obtained numerically. Moreover, starting from a certain value
of $d$, one has to take principal values of integrals in order to
find the FP in the frames of the resummation procedure considered
here. A similar problem  was observed within two-loop
approximation in the massive scheme \cite{Dudka01a} already at
$d=3$. Furthermore, within the five loop order  of the massive
scheme \cite{new_Ital} no evidence for this new stable FP was
found. Therefore, we consider this new FP as an unphysical one.

\subsubsection{Cubic distribution}\label{III.2}
Imposing the minimal subtraction renormalization conditions
(\ref{mincond}) to the vertex functions of the theory with the
effective Hamiltonian (\ref{cubham}) we get the following RG
functions in the two-loop approximation:
\begin{eqnarray}\label{betacub1}
\beta_u\!&=&\!-\varepsilon u{+}\frac
{mn{+}8}{6}{u}^{2}{+}\frac{m+2}{3}vu
{+}\frac{n+2}{3}uw{+}yu{+}\frac{1}{3}wv{-} \nonumber\\
&&\frac{3mn+14}{12}{u}^{3}{-}{\frac {11\left
(m+2\right)}{18}}v{u}^{2} {-}{\frac {5\left (m+2\right
)}{36}}{v}^{2}u {- }\nonumber\\ &&{\frac {11\left (n+2\right
)}{18}u^{2}w}{-} {\frac {5\left (n+2\right
)}{36}}u{w}^{2}{-}{\frac {11}{6}}{u}^{2}y-{\frac {5}{12}}{y}^{2}u
{-}\nonumber\\ &&\frac{3}{2}vuw-\frac{5}{6}wyu-\frac{5}{6}vuy
-\frac{1}{9}{v}^{2}w-\frac{1}{9}{w}^{2}v,
\end{eqnarray}

\begin{eqnarray}\label{betacub2}
\beta_v\!&=&\!-\varepsilon v{+}\frac
{m+8}{6}{v}^{2}{+}2vu{+}\frac{2}{3}wv{+}yv {-}\nonumber\\ &&
\frac{3m+14}{12}{v}^{3}{-}\! {\frac {11m{+}58}{18}}{v}^{2}u
{-}{\frac {5mn{+}82}{36}}v{u}^{2} {-}{\frac {n{+}10}{36}}{w}^{2}v
{-}\nonumber\\ &&\!\frac{5}{6}vwy{-}{\frac
{31}{18}}{v}^{2}w{-}{\frac {11}{6}}{v}^{2}y {-}{\frac
{5}{12}}{y}^{2}v{-}{\frac {17}{6}}vuy{-}{\frac {5n{+}34}{18}}wvu,
\end{eqnarray}

\begin{eqnarray}\label{betacub3}
\beta_w\!&=&\!-\varepsilon w{+}\frac
{n+8}{6}{w}^{2}{+}2wu{+}\frac{2}{3}wv{+}yw {-}
\nonumber\\&&\frac{3n+14}{12}{w}^{3}{-} {\frac
{11n{+}58}{18}}{w}^{2}u{-}{\frac {5mn{+}82}{36}}w{u}^{2}
{-}\nonumber\\ &&{\frac {31}{18}}{w}^{2}v{-}{\frac
{m{+}10}{36}}{v}^{2}w {-}{\frac {11}{6}}{w}^{2}y {-}{\frac
{5}{12}}{y}^{2}w{-}\nonumber\\ &&{\frac {17}{6}}wuy{-}{\frac
{5m+34}{18}}wvu{-}\frac{5}{6}vwy,
\end{eqnarray}

\begin{eqnarray}\label{betacub4}
\beta_y&=&-\varepsilon
y+\frac{3}{2}{y}^{2}+2yu+2yv+2wy+\frac{4}{3}wv{-} \nonumber\\
&&{\frac {17}{12}}{y}^{3}- {\frac {5mn+82}{36}}{u}^{2}y{-}{\frac
{23}{6}}{y}^{2}u{-}{\frac {23}{6}}{y}^{2}v {-}{\frac
{23}{6}}{y}^{2}w{-}\nonumber\\ &&{\frac {5m+82}{36}}{v}^{2}y{-}{
\frac {5n+82}{36}}{w}^{2}y{-}\frac {n{+}18}{9}{w}^{2}v{-} \frac
{m+18}{9}{v}^{2}w{-}\nonumber\\&&{\frac {41}{6}}vwy{ -}{\frac
{5n+82}{18}}wyu{-}{\frac {5m+82}{18}}vuy{-}\frac{8}{3}vuw,
\end{eqnarray}

\begin{eqnarray}\label{gammacub1}
\gamma_\phi&=&0+\nonumber\\ &&{\frac
{mn+2}{72}}{u}^{2}{+}\frac{{y}^{2}}{24}{+}{\frac {n+2}{72}}{w}^{2}
{+}{\frac {m+2}{72}}{v}^{2}{+}\frac{yu}{12}{+}\nonumber\\
&&\frac{wv}{12}{+}\frac{yv}{12}{+}\frac{m+2}{36} vu{+}\frac
{n+2}{36}wu{+}\frac{wy}{12}
\end{eqnarray}

\begin{eqnarray}\label{gammacub2}
\bar{\gamma}_\phi^2&{=}&\frac{mn+2}{6}u{+} \frac {m+2}{6}v{+}\frac
{n+2}{6}w+\frac {y}{2}{-}\nonumber\\ &&\frac {mn+2}{12}{u}^{2}{-}
\frac {m+2}{12}{v}^{2}{-}\frac{n+2}{12}{w}^{2}{-}\frac
{{y}^{2}}{4} {-}\nonumber\\ &&\frac {m+2}{6}vu{-}\frac
{n+2}{6}wu{-}\frac{yu}{2}{-}\frac{wv}{2}{-}\frac{yv}{2}{-}\frac{wy}{2}.
\end{eqnarray}
Here, $u,v,y,y$ stand for the renormalized couplings and again the
replica limit $n=0$ will be considered.

As noted in the previous subsection \ref{III.1}, two independent
ways to analyse expressions (46)-(51) are possible. The
$\varepsilon$-expansion to first order in $\varepsilon$ was
performed in Ref. \cite{Aharony75}. It reported  14 FPs but no
stable and accessible one was found among them. However,
subsequently Mukamel and Grinstein \cite{Mukamel82} pointed to the
possibility of a second order phase transition with exponents of
the weakly diluted quenched Ising model \cite{Grinstein76}.
Indeed, performing perturbation theory expansion to order
$\varepsilon^2$ not only the corrections to the existing FP
 are obtained
 but  new FPs \cite{Dudka01c,Dudka02,Dudka01b} appear.
  Their appearance  is caused by the
degeneracy of $\beta$-functions   at the one loop level. In
particular, in the replica limit the one-loop parts (the first
lines respectively) of $\beta_w$, $\beta_y$ (\ref{betacub3}),
(\ref{betacub4}) and $\beta_u$, $\beta_y$ (\ref{betacub1}),
(\ref{betacub4}) are degenerated for any $m$ at $u=v=0$ and at
$w=v=0$, correspondingly. Another one-loop degeneracy of the
$\beta_u$, $\beta_v$, $\beta_y$ functions with $w=0$ occurs at
$m=2$. The last degeneracy was observed also for the diluted cubic
model \cite{cubdil,Shpot}. This implies a
$\sqrt\varepsilon$-expansion \cite{Grinstein76,sqrt} for the FPs
\cite{Dudka01c,Dudka02,Dudka01b} rather than an
$\varepsilon$-expansion. Among the FPs found with the help of  the
$\sqrt\varepsilon$-expansion, the FP with coordinates
$w^*<0,\,y^*>0$, $u^*=v^*=0$ is stable. Moreover, this point is
reachable from the initial values of the couplings. As far as it
is the FP of the diluted Ising model one concludes, that in the
critical region the RAM with a {\em cubic distribution} of random
anisotropy axes (\ref{cubham}) decouples into $m$ independent
diluted Ising models and the phase transition is governed by the
familiar random Ising model critical exponents.

This picture found its further confirmation by the fixed $d$
approach in the two-loop \cite{Dudka01c,Dudka02,Dudka01b} and
later in the five-loop \cite{new_Ital} approximation of the
massive scheme. The minimal subtraction scheme  corroborates these
results. Applying the resummation procedure (\ref{r1})-(\ref{r3})
to the $\beta$-functions (\ref{betacub1})-(\ref{betacub4})  we get
 16 FPs. In Table~\ref{tabmin2} we present the numerical
values of the FPs coordinates with $u^*>0$, $v^*>0$, $w^*<0$. The
numeration of FPs  correspond to that of Refs.
\cite{Aharony75,Dudka01c,Dudka02,Dudka01b}. Only FP XV and the
``polymer" FP III are stable. However the last one is not
reachable from physical initial conditions.  FP XV in
Table~\ref{tabmin2}  corresponds to the stable FP found in the
$\sqrt\varepsilon$-expansion \cite{Dudka01b}, and has coordinates
with $u^*=v^*=0$, $w^*<0$ and $y^*>0$, thus it is accessible  from
typical initial values for the  couplings.

\begin{table}
\begin{center}
 {\begin{tabular}{|c|c|c|c|c|c|c|c|} \hline
FP&$m$&$u^*$&$v^*$&$w^*$&$y^*$&$\nu$&$\gamma$\\
 \hline I &$\forall m$& 0& 0& 0& 0&1/2 &1 \\
\hline &2& 0 & 1.1415 & 0 & 0 & 0.684 &1.344 \\ II&3&0 & 1.0016 &
0 & 0 & 0.720 & 1.411\\ &4& 0 & 0.8878 & 0 & 0 & 0.751 & 1.470\\
\hline III&$\forall m$& 1.5281 & 0 & 0 & 0 & 0.599 & 1.185 \\
\hline V&$\forall m$&0 & 0 & 0 & 1.3146 & 0.644 & 1.269\\ \hline
&2&0.1429 & 0.9923 & 0 & 0 & 0.665 & 1.307\\ VI&3&0.3411 & 0.6965
& 0 & 0 &0.660&1.298\\ &4&0.4546 & 0.5323 & 0 & 0 & 0.656 &
1.291\\ \hline VII&$\forall m$&2.5382 & 0 & -2.5382 & 0 &1/2 &1 \\
\hline &2&0 & 1.9719 & 0 & -1.3146 & 0.644&1.269\\ VIII&3&0 &
0.8568 & 0 & 0.2270 &0.717&1.405\\ &4&0 & 0.4435 & 0 & 0.7555 &
0.715 & 1.403\\ \hline &2&-0.2506 & 2.4494 & 0 & -1.6330 & 0.667 &
1.312\\ IX&3&0.3404 & 0.7275 & 0 & -0.0511&0.660&1.298
\\ &4&0.4325 & 0.4342 & 0 & 0.2149 & 0.657&1.292\\ \hline
X&$\forall m$&0.7311 & 0 & -0.7311 & 1.3146 &0.644&1.269\\ \hline
XV&$\forall m$&0 & 0 & -0.2506 & 1.6330 & 0.667 & 1.312\\ \hline
\end{tabular}}
\caption{\label{tabmin2} Fixed points coordinates and critical
exponents $\nu$ and $\gamma$ as obtained on the base of resummed
two-loop $\beta$- and  $\gamma$-functions for the RAM with {\em
cubic distribution} of random anisotropic axes. FP III and FP XV
are stable. Note that the exponents may take the same values for
different FPs. }
\end{center}
\end{table}

Applying the resummation procedure (\ref{r1})--(\ref{r3}) to the
$\beta$-functions (\ref{betacub1})-(\ref{betacub4}) we have not
found any other stable FPs in the region of interest. Thus one
concludes that the effective Hamiltonian (\ref{cubdist}) in the
asymptotic region
  reduces to a product of $m$ effective
Hamiltonians of a weakly diluted quenched random site Ising model.
This means that independent on the dimension $m$ of the order
parameter field in (\ref{ffrepresent}) the system is characterized
by the same set of critical exponents which are those of the
weakly diluted random site quenched Ising model.

It is important to note that the FP XV does not describe critical
behaviour if the random anisotropy axes distribution differs from
the cubic one. In such a case the effective Hamiltonian
(\ref{fulham}) has to be considered and FP XV turns out to be
unstable with respect to the term with $z_0$. It was stated
already in the one-loop study \cite{Mukamel82} and checked by
nonperturbative arguments \cite{new_Ital}. However critical
properties of Hamiltonian (\ref{fulham}) were studied only in
one-loop order so far.

From the stability exponents of the stable  FP XV,  one can obtain
information about how fast the couplings approach their asymptotic
values when one goes to $T_c$. Approaching the critical point the
correction to scaling enters the power-law  behaviour of
thermodynamical quantities. This correction is defined by the
smallest stability exponent of the stable accessible FP. We get
the following values for the stability exponents of the FP XV:
$\omega_1\approx0.1935$, $\omega_2\approx0.4939$,
$\omega_u\approx0.0036$, $\omega_v\approx0.0036$. Here, $\omega_1$
and $\omega_2$ are the familiar stability exponents of the
random-site Ising model. Currently, accurate values of these
exponents within five-loop \cite{Folk00} and six-loop
\cite{Pelissetto00,Pelissetto} approximations are known. Let us
note, that the leading correction to scaling exponents for the
model (\ref{cubham}) is determined by the stability exponents
$\omega_u$, $\omega_v$. { General arguments \cite{new_Ital} state
that the leading correction exponent can be expressed by
$\omega=\omega_u=\omega_v=-\alpha_r/\nu_r$, where $\alpha_r$ and
$\nu_r$ are the random-site Ising model exponent of the specific
heat and the correlation length respectively. Indeed, this is
fulfilled: $\alpha_r/\nu_r\approx -0.0036$, in two loop order. The
Wegner correction to scaling $\Delta=\omega\nu_r=-\alpha_r$ is
equal to the random-site Ising model specific heat critical
exponent with opposite sign \cite{new_Ital}, which has been
calculated with higher accuracy up to six loops
\cite{Pelissetto02,Folk00,Calabrese03} and is estimated as
$\Delta\approx 0.049\pm0.009$ \cite{new_Ital}.}

For the other FPs, we recover the familiar two-loop numerical
results for the Gaussian (FPs I, VII), the $m$--vector (FP II),
the polymer $O(n=0)$ (FP III), the Ising (FPs V, X), the diluted
$m$--vector (FP VI),  and cubic (FP VIII) models. FP IX belongs to
a new universality class. In the Table \ref{tabmin2} we give the
numerical values of the critical exponents in these FPs as well:
if the flow from the initial values of couplings pass near one of
these FPs one may observe an effective critical behaviour governed
by the values of the corresponding  critical exponents. This will
be discussed in the next subsection in more detail.

\subsection{Effective critical behaviour}\label{III.III}

It is hard to approach the critical point close enough to observe
asymptotic critical behaviour in the experiments as well as in the
computer simulations. Therefore, very often one deals with the
{\em nonasymptotic} critical behaviour described by the {\em
effective critical exponents}. These exponents are defined as
logarithmic derivatives of corresponding thermodynamic quantities
with respect to the reduced temperature $\tau$ \cite{Kouvel64}.
Within the RG approach the effective critical exponents are
defined in the region, where the scale dependent couplings
$u_i(\ell)$ have not reached their FP values and depend  on the
flow parameter $\ell$. In particular for the susceptibility
critical exponent $\gamma_{\rm eff}$ it results in the expression:
\begin{eqnarray}
 \label{effgamma}\gamma^{-1}_{\rm eff}(\ell)&{=}&1-
\frac{\bar{\gamma}_{\phi^2}(\{u_i(\ell)\})}{2-{\gamma}_{\phi}(\{u_i(\ell)\})}+\dots,
\end{eqnarray}
where the dots  mean contributions from amplitude parts of the
corresponding thermodynamic function. In our calculations we
neglect these parts assuming contribution of the amplitude
functions to the crossover to be small \cite{note2}.

If the system of differential equations (\ref{floweq}) has a
stable FP $\{u^*_i\}$ accessible from the initial conditions, this
FP corresponds to a critical point of the system. Then RG flows in
the limit $\ell\to0$ reach this FP, and the effective critical
exponents in this limit coincide with their asymptotic
counterparts. However the initial conditions being near the origin
in the space of couplings (i. e. near the Gaussian FP) the flow
may approach unstable FPs on its way to the stable one. Then the
effective exponents may take values corresponding to these
unstable  FPs.

In the next subsections the effective critical behaviour of the
both  Hamiltonians (\ref{isham}) and (\ref{cubham}) is
investigated in the two-loop approximations of the minimal
subtraction scheme. As the study  of the diluted Heisenberg
magnets \cite{Dudka03} shows, the two-loop approximation is quite
precise to describe the experimentally observed behaviour of the
effective critical exponents. Moreover, the investigations of
effective critical behaviour of the random-site Ising model within
high-loop order approximations \cite{Pelissetto} in a
semiquantative way fully agree with the qualitative picture
obtained in the lower approximations \cite{Folk00}. Therefore
since the effective critical exponents are nonuniversal
quantities, we consider the two-loop approximation as satisfactory
approach for their description.

\subsubsection{Isotropic distribution}

The solutions of the system of flow equations (\ref{floweq}), with
the $\beta$-functions given by the resummed expressions
(\ref{betais1})-(\ref{betais3}) determine the scale-dependent
couplings for the RAM with the random anisotropy axes {\em
isotropic distribution}. The initial values for differential
equations usually are chosen in the vicinity Gaussian FP
describing the background behaviour of system. For the RAM these
values are restricted by the subspace $u>0$, $v>0$ and $w<0$.
There is also an additional condition for the bare couplings:
$w_0/u_0=-m$. However, starting from renormalized coupling values
that satisfy this condition, the flows fast run into the region
with large negative $w$, where poles appear in the integrals for
the resummed functions (\ref{r3}). In addition, the existence of
such a region questions the  stability of the Hamiltonian ({see
Section \ref{II}). Therefore, we enlarge the region of the initial
values of couplings, not necessarily restricting them to the
condition $w/u=-m$. Since the flow picture for the case $m=2$ can
be analysed without appealing to calculation of principal values
of integrals in the resummed RG functions and thus is much more
reliable, we have chosen this case to be presented here.

\begin{figure}
\includegraphics[width=0.7\textwidth]{{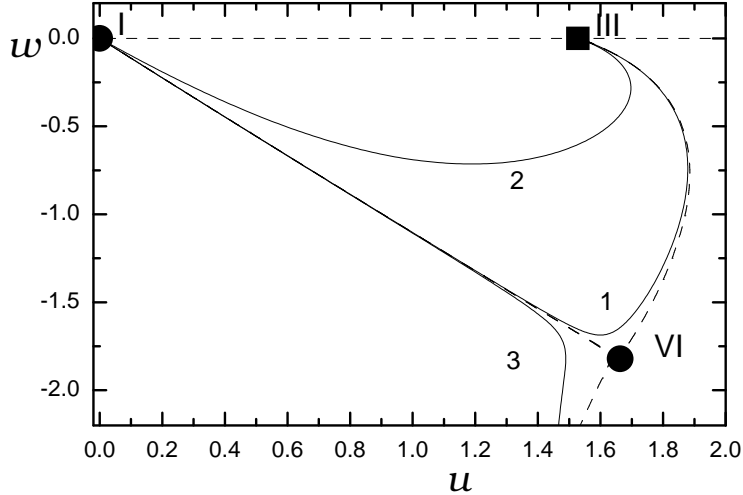}}
\caption{\label{bad} Flows in the space of couplings at $v=0$,
$m=2$. A square denotes the stable FP, circles denote unstable FPs
and dashed lines show separatrices connecting two FPs. Curves 1
and 2 correspond to a flow for the ratio $|w/u|$ chosen close
 above the separatrix between the FPs I and VI, while the flow
3 starts from a value $|w/u|$ below the separatrix.}
\end{figure}

\begin{figure}
\includegraphics[width=0.7\textwidth]{{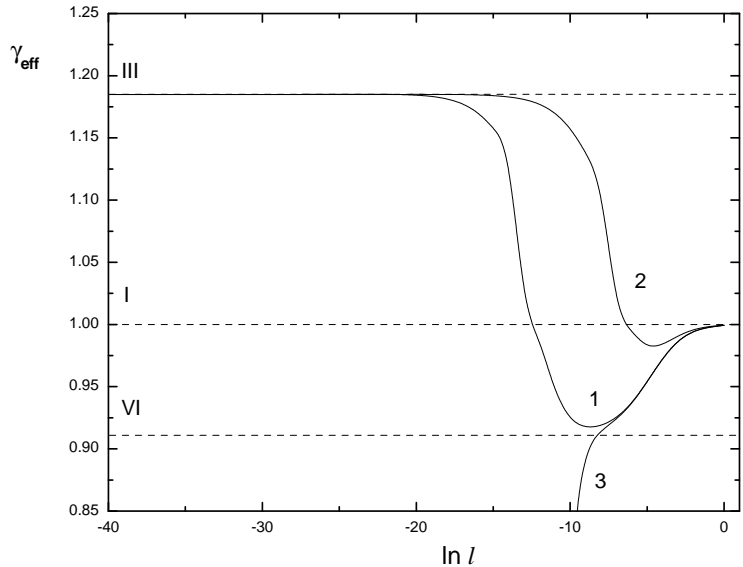}}
\caption{\label{badgam} Effective critical exponent $\gamma_{\rm
eff}$ as a function of the logarithm of the flow parameter, $m=2$.
Curves correspond to flows of Fig.~\protect\ref{bad} (denoted by
the corresponding numbers). The dashed lines indicate the values
of $\gamma$ at the corresponding FP.}
\end{figure}

In the Fig.~\ref{bad} we show the flows that start from the
initial values with $v=0$.  All flows starting from values of
couplings  below the separatrix connecting FP I and VI  run into
the region of large negative $w$, as discussed above (e. g. flow 3
in the Fig.~\ref{bad}). In fact they are ``runaway" solutions
which may indicate the possibility of a first order transition.
The curves 1 and 2 in the Fig.~\ref{bad} are obtained from initial
values located above the separatrix connecting FPs I and VI. We
expect that these flows can not be observed for the RAM because
they
 are far away from the condition $w/u=-m$.
The flows  starting above the separatrix connecting FP I with FP
VI are influenced by the unstable FP VI
 and finally attracted by the ``polymer" stable FP III.
Fig.~\ref{badgam} shows the curves of magnetic susceptibility
effective critical exponent $\gamma_{\rm eff}$ corresponding to
the flows of the Fig. \ref{bad}. We do not present results for
other effective critical exponents, since they demonstrate
qualitatively the same behaviour as $\gamma_{\rm eff}$.

Let us now consider the possible scenarios for the  effective
critical exponents if all  initial values of couplings are
non-zero. Flows starting from physical initial values for the RAM
 are either attracted by the stable FP we rejected as unphysical
 (see the last paragraph of subsection \ref{III.1}) or have runaway
character.  However the critical behaviour in the initial phase of
the flow can be influenced by the physical unstable FPs. At
initial values appropriate to the RAM a flow can be realized, that
is affected by the unstable FP II, which describes critical
behaviour of pure $m$-vector model. Thus the effective critical
exponent corresponding to this flow (Fig.~\ref{isotrgav})
demonstrates regions with the critical exponent of the $m$-vector
model (in the given case for $m=2$). Monte Carlo calculations for
$m=2$ report critical exponents similar to the pure XY model
\cite{Rossler}. Our theory suggests that the exponents found are
effective ones.

\begin{figure}
\includegraphics[width=0.7\textwidth]{{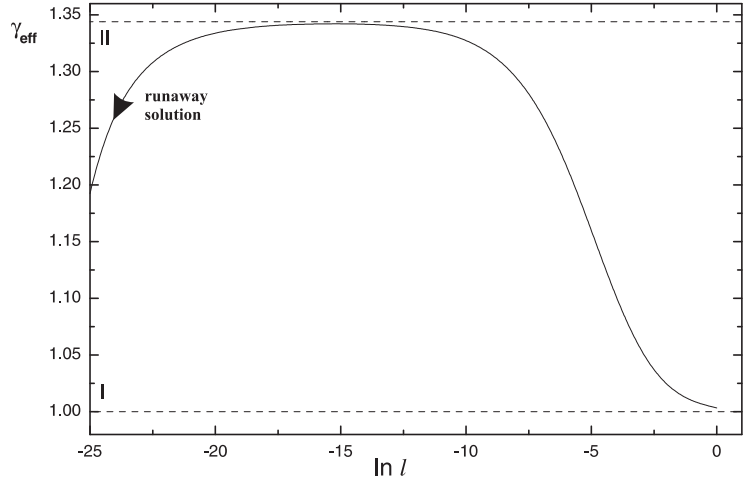}}
\caption{\label{isotrgav} The effective critical exponent
$\gamma_{\rm eff}$  as a function of the logarithm of the flow
parameter for isotropic distribution of the local anisotropy axes,
$m=2$. Dashed lines indicate the value of $\gamma$ at the
corresponding FPs.}
\end{figure}

\subsubsection{Cubic distribution}

Flows for the cubic distribution of the random anisotropy axes
 have been considered already within the five-loop
massive scheme \cite{new_Ital}, however the behaviour of effective
critical exponents has not been studied. They confirm
 that the FPs of the random Ising FP XV and ``polymer" FP III
models are the only stable FPs. Moreover, the flows from initial
conditions satisfying $u>0$, $w<0$, $u+w<0$ and $v+y>0$, $mv+y>0$
according to (\ref{st_cond}) are never  attracted by the
``polymer" FP III. Therefore in our investigation we restrict the
initial conditions to this region. Starting with large negative
$w$, the flow fast runs into the region, where the resummation
method does not work. Thus the initial conditions are chosen close
to the line $w/u=-m$.
\begin{figure}
\includegraphics[width=0.7\textwidth]{{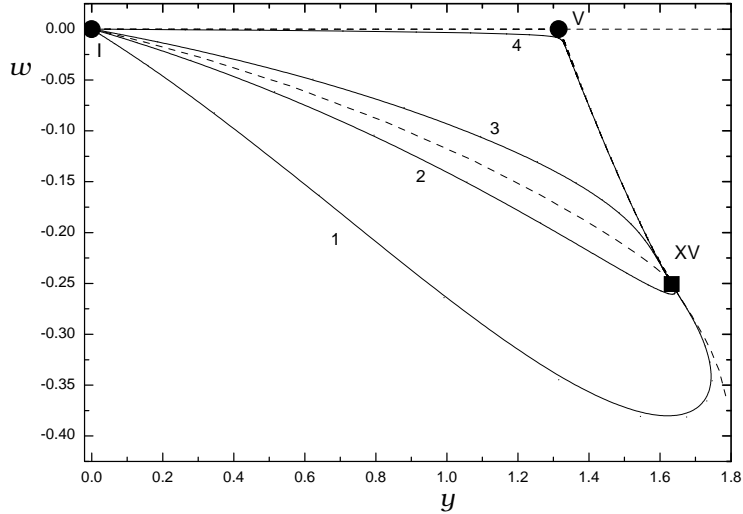}}
\caption{\label{cubic} Flows in  space of couplings for $u=v=0$.
The square denotes the stable FP, the circles mean unstable FPs
and the dashed lines are separatrices connecting two FPs. }
\end{figure}

\begin{figure}
\includegraphics[width=0.7\textwidth]{{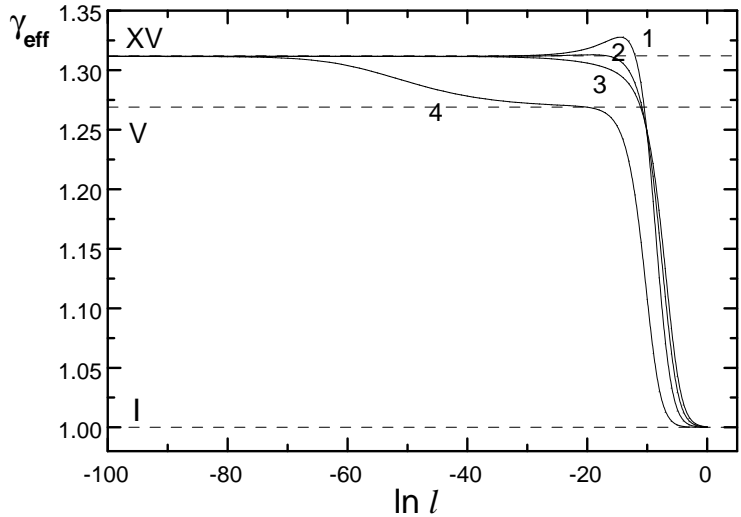}}
\caption{\label{cubicg} Effective critical exponent $\gamma_{\rm
eff}$ as a function of the logarithm of the flow parameter. Curves
correspond to flows of Fig.~\protect\ref{cubic} denoted by
corresponding numbers. Dashed lines indicate the value of $\gamma$
at the corresponding FPs.}
\end{figure}

As an illustration, let us first consider the trajectories with
the initial conditions $u=v=0$. Solutions of the flow equations in
this case
  are plotted
in the Fig.~\ref{cubic}. The corresponding  scenarios of the
behaviour of the effective critical exponent $\gamma_{\rm eff}$
are shown in the Fig.~\ref{cubicg}. The particular features of
initial conditions with $u=v=0$ is, that the effective critical
exponents are independent of the order parameter dimension $m$
since all $m$-dependent terms in the flow equations drop out. At
large values of ratio $|w/y|$ one can observe flows which seem to
be not attracted by the stable FP XV at the beginning, but finally
the flows end in this FP. Therefore effective critical exponents
with values overshooting the asymptotic ones can be observed
(curve 1 in the Fig.~\ref{cubicg}). If flows are strongly affected
by the unstable FP V, then before they crossover to asymptotic
values of the dilute Ising model the critical exponents of the
pure Ising model (corresponding to FP V) are observed (curve 4 in
the Fig.~\ref{cubicg}). Since the effective Hamiltonian
(\ref{cubham}) with $u_0=v_0=0$ corresponds to the diluted Ising
model, in  Figs.~\ref{cubic} and \ref{cubicg} the picture obtained
for that model \cite{Folk00,Pelissetto} is recovered.
\begin{figure}
\includegraphics[width=0.65\textwidth]{{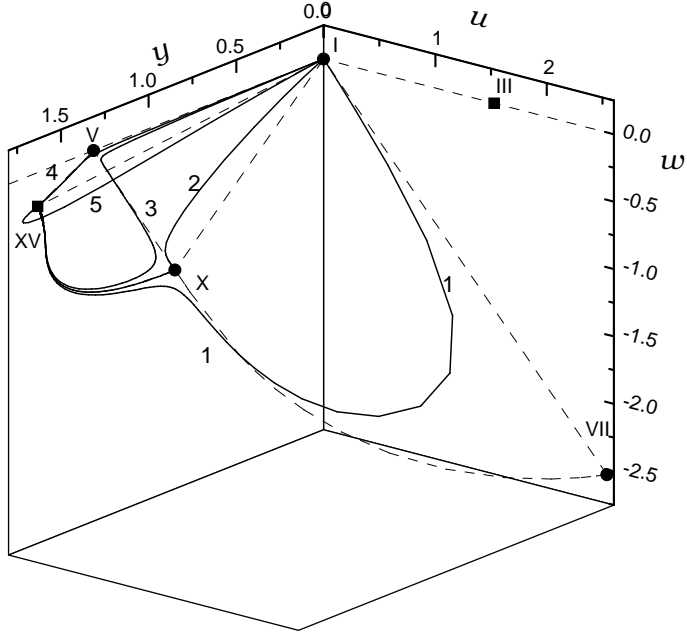}}
\caption{\label{cubic3} Flows in  space of couplings for $v=0$.
The squares denote stable FPs, the circles mean unstable FPs and
the dashed lines are separatrices connecting two FPs. }
\end{figure}

\begin{figure}
\includegraphics[width=0.7\textwidth]{{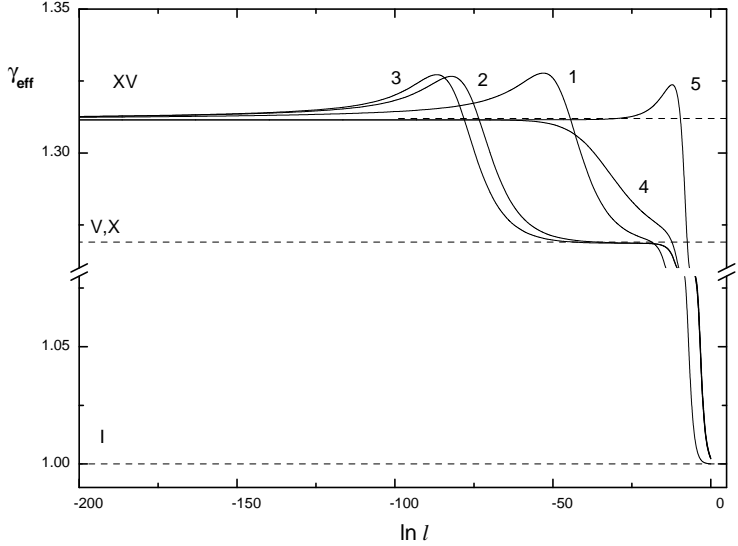}}
\caption{\label{cubicgam} Effective critical exponent $\gamma_{\rm
eff}$ as a function of the logarithm of the flow parameter. Curves
correspond to flows of Fig.~\protect\ref{cubic3} denoted by
corresponding numbers. Dashed lines indicate the value of $\gamma$
at the corresponding FPs.}
\end{figure}

If we ``switch on" the coupling  $u_0$ (but leave $v_0=0$) a
richer scenario of flows is obtained. The typical ones
 are given in the Fig.~\ref{cubic3}. Also in this case
results are independent on the order parameter dimension $m$,
since all $m$-dependent terms drop out of the flow equation. The
effective critical exponent $\gamma_{\rm eff}$ corresponding to
these flows is shown in the next Fig.~\ref{cubicgam}. There are
two FPs V and X that correspond to the pure Ising model.  So,
there exists the possibility that flows starting from certain
initial values in the space of couplings first are affected by the
unstable FP V of the pure Ising model, then are attracted by the
Ising FP X and only then they approach the stable FP XV. It means
that within a wide range of temperature the critical exponents of
pure Ising model may be observed before approaching the asymptotic
regime (curve 3 in the Fig.~\ref{cubicgam}). A characteristic
feature of almost all flows is the non-monotonic crossover  to the
asymptotic values.

The effective critical behaviour, which can be observed if all
initial couplings are non-zero,  depends on $m$ and is described
by Figs.~\ref{cub2},~\ref{cub3} and~\ref{cub4} for values of spin
component $m=2,\,3,\,4$ correspondingly. Apart from the
characteristic behaviour of effective critical exponent
$\gamma_{\rm eff}$ presented by curves 1-5 of Fig.~\ref{cubicg}, a
new scenarios of effective critical behaviour appear. They are
presented by curves 6 and 7 in the Figs. \ref{cub2}-\ref{cub4},
that demonstrate regions with effective critical exponent of the
pure $m$-vector model (FP II) and the cubic model (FP VIII)
correspondingly.

In connection with this result, one should mention Monte Carlo
simulations of the $d=3$ RAM with  {\em cubic distribution} of the
random anisotropy axes \cite{Fisch93}. For $m=2$ and weak
anisotropy the results are consistent with a second order phase
transition with critical exponents, which are close to those of
the pure XY model. The theory shows that the asymptotic critical
behaviour is governed by the random Ising critical exponent, FP
XV, while critical exponents of pure model, FP II, are possible as
effective ones in the certain temperature region.

\begin{figure}
\includegraphics[width=0.7\textwidth]{{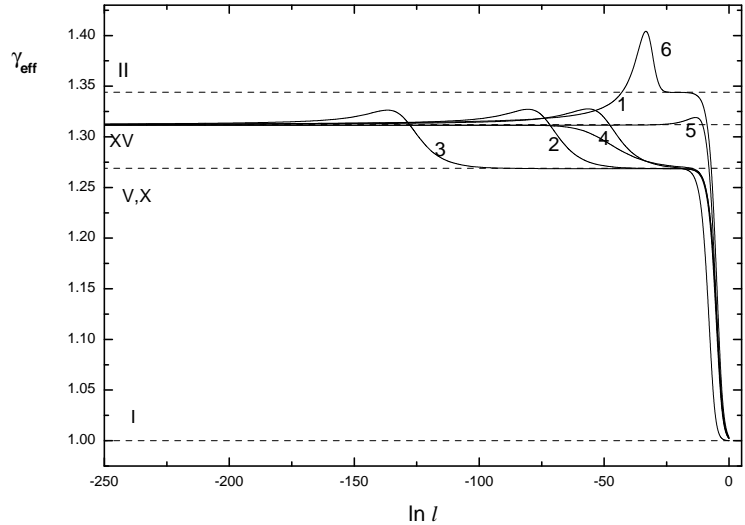}}
\caption{\label{cub2} Effective critical exponent $\gamma_{\rm
eff}$ as a function of the logarithm of the flow parameter for
order parameter dimension $m=2$. Dashed lines indicate the value
of $\gamma$ at the corresponding FPs.}
\end{figure}

\begin{figure}
\includegraphics[width=0.7\textwidth]{{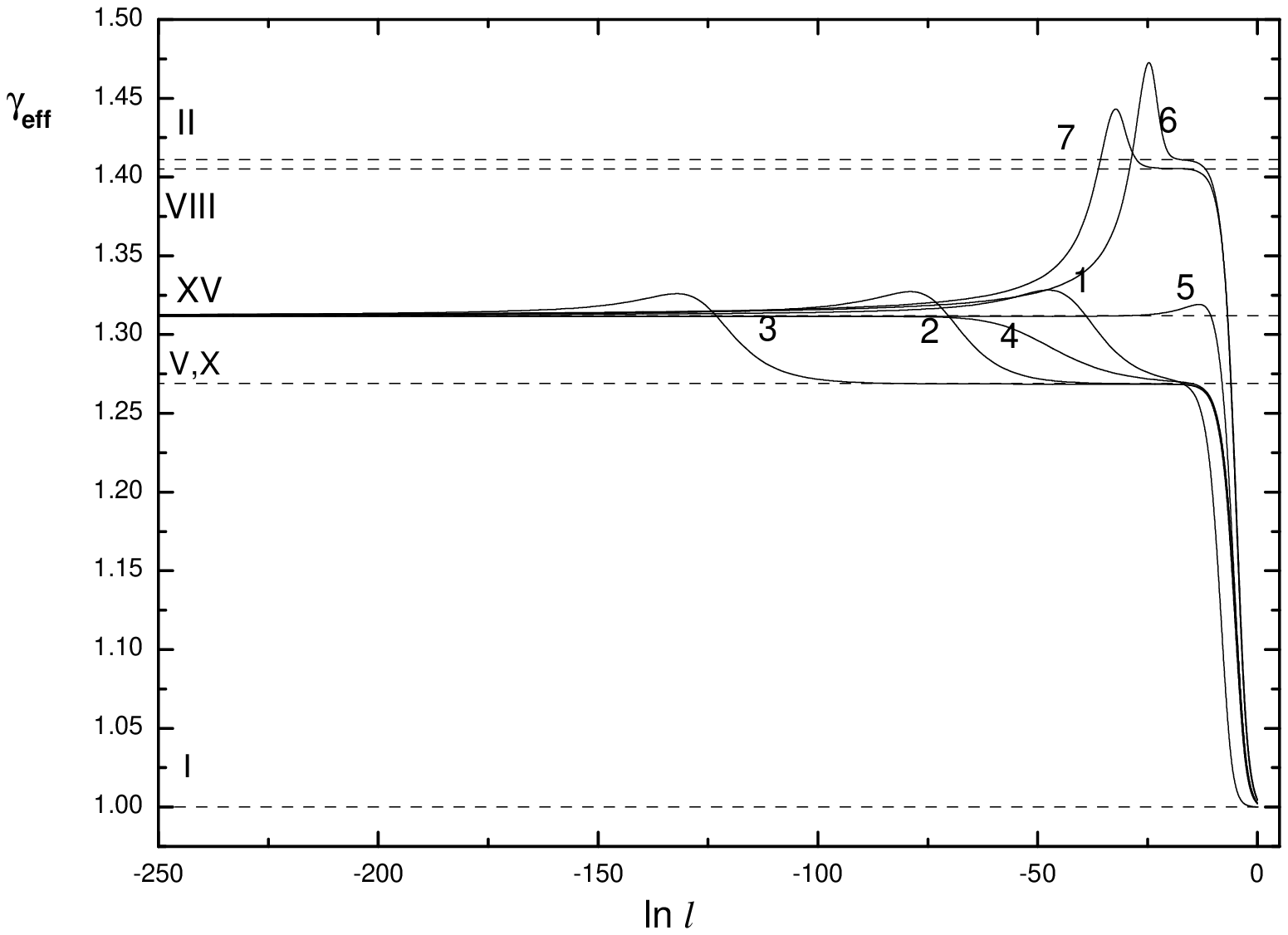}}
\caption{\label{cub3} Effective critical exponent $\gamma_{\rm
eff}$ as a function of the logarithm of the flow parameter for
order parameter dimension $m=3$. Dashed lines indicate the value
of $\gamma$ at the corresponding FPs.}
\end{figure}

\begin{figure}
\includegraphics[width=0.7\textwidth]{{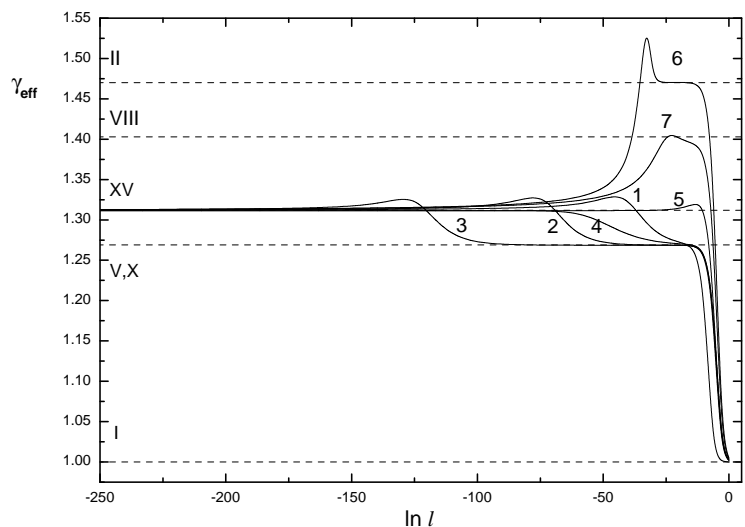}}
\caption{\label{cub4} Effective critical exponent $\gamma_{\rm
eff}$ as a function of the logarithm of the flow parameter for
order parameter dimension $m=4$. Dashed lines indicate the value
of $\gamma$ at the corresponding FPs.}
\end{figure}

\section{Conclusions}\label{V}

In the present paper we have studied the critical behaviour of
three dimensional random anisotropy magnets. The experimental,
numerical and theoretical investigations performed so far give
contradictory results. However, two main conclusions can be drawn
from the existing data. The random anisotropy with {\em isotropic
distribution} of quenched anisotropy axes destroys the second
order phase transition of the pure system, while for the {\em
anisotropic distribution} there exists the possibility of a second
order phase transition.

The standard tool for describing the critical behaviour of
different systems, the field-theoretical RG, gives results in
consistency with these conclusions. In particular, the absence of
a stable accessible FP for an {\em isotropic distribution} of the
random axes means an absence of a second order phase transition.
First derived on the basis of a one-loop RG calculation
\cite{Aharony75}, this conclusion found its further support in the
analysis of the two- \cite{Dudka01a,Dudka01c,Dudka02} and the
five-loop \cite{new_Ital} massive RG functions refined by
resummation.

The two-loop approximation within the massive RG scheme applied to
the case, when random axes are distributed along the edges of the
$m$-dimensional hypercube, lead to an answer about the second
order phase transition with the same set of critical exponents as
those of the weakly diluted Ising systems
\cite{Dudka01c,Dudka02,Dudka01b}. This was confirmed in a recent
five-loop RG calculations \cite{new_Ital}.

Although the majority of the experimental and simulation data show
a effective critical behaviour, the effective critical behaviour
of random anisotropy magnets was not studied extensively so far.
We used the minimal subtraction scheme of the field theoretical RG
approach to analyse such a behaviour. The calculations were
performed within two-loop approximation combined with resummation
of perturbation theory expansions. The results concerning the
asymptotic critical behaviour confirm the picture obtained in the
massive scheme \cite{Dudka01a,Dudka01c,Dudka02,Dudka01b,new_Ital}.

Studying nonasymptotic critical behaviour of the $d=3$ RAM with
{\em isotropic} as well as with {\em cubic distributions} of
anisotropy axes we calculated the effective critical exponents.
The results show that for certain initial conditions the effective
critical behaviour of the random anisotropy magnets with {\it
isotropic distribution} of random axes can be governed by critical
exponent of the $m$-vector model at least for $m=2$ in a certain
temperature interval. Also we investigated effective critical
behaviour of systems described by Hamiltonian (\ref{isham}) with
values of bare couplings differing from those of the RAM.

In the case of the RAM with {\it cubic distribution} of anisotropy
axes, various scenarios of effective critical behaviour can be
observed. The crossover from background behaviour described by the
Gaussian critical exponents to the asymptotic random Ising
behaviour demonstrates the existence of temperature regions where
effective critical exponents coincide with values of the critical
exponents of pure Ising, cubic and $m$-vector  models. A
characteristic feature of the majority of  the effective critical
exponent  is the appearance of peaks before the crossover to the
asymptotic limit.

In conclusion, we want to attract attention to the importance of
determining the {\em nonasymptotic} critical regime in the
investigation of the critical behaviour of random anisotropy
magnets. As our investigation shows being  not  close enough  to
the {\em asymptotic} critical regime different values of critical
exponents can be obtained, they have been shown to be {\em
effective} ones.

This work was supported by the Fonds zur F\"orderung der
wissenschaftlichen Forschung under Project No. 16574-PHY. Yu. H.
and M. D.  acknowledge French-Ukrainian cooperation Dnipro project
and useful comments of Bertrand Berche and Taras Yavors'kii.

\appendix
\section{Appendix}

  Here, we describe the procedure of the RAM representation in terms of
the functional integrals. The case of weak anisotropy is of
special interest and to introduce the controlling parameter
$D/\bar{J}$ an appropriate normalization of spins can be used
$\vec{S}'=\vec{S}\sqrt{\bar J}$. Then, Hamiltonian (\ref{origham})
is rewritten in form:
\begin{equation}\label{ham}
{\mathcal H} =  - \sum_{{\bf R},{\bf R'}} \frac{J_{{\bf R},{\bf
R'}}}{\bar{J}} \vec{S'}_{\bf R} \vec{S'}_{\bf R'}
-\frac{D}{\bar{J}}\sum_{{\bf R}} (\hat {x}_{\bf R}\vec{S'}_{\bf
R})^{2}.
\end{equation}
It is useful to perform a Fourier transformation for the
interaction part of (2) and leave the random axis part as it is,
after that the Hamiltonian reads:
\begin{equation}\label{hamram}
{\mathcal H} =  - \sum_k\nu(k)\vec{S'}_{\bf k}\vec{S'}_{\bf -k}-
D'\sum_{R}(\hat x_{\bf R}\vec{S'}_{\bf R})^2,
\end{equation}
where $\vec{S'}_{\bf k}$ and $\nu(k)$ are Fourier transforms of
spin and interaction correspondingly, $D'=D/\bar{J}$.

 For the fixed configuration of the local
random anisotropy axes $\{\hat x_{\bf R}\}$ the
configuration-dependent partition function is written
\begin{equation}\label{partfun}
Z(\{\hat x_{\bf R}\})={\rm Sp}\,\,e^{-{\beta\mathcal H}},
\end{equation}
where the Hamiltonian $\mathcal H$ is given by (\ref{hamram}),
$\beta$ is the inverse temperature and the trace means integration
over the surface of $m$-dimensional hypersphere with unit radius:
${\rm Sp}(\dots)=\int d\vec{S'}_{\bf
R}(\dots)\delta(|\vec{S'}_{\bf R}|-1)$.

To avoid taking trace of products of spins one  can apply the
Stratonovich-Hubbard transformation to  expression (\ref{partfun})
and arrive at the  following form of the partition function:
\begin{eqnarray}\label{represent}
Z(\{\hat x_{\bf R}\})&\sim&{\rm Sp} \int (d\vec{\phi})\exp
\Big\{-\frac{1}{4}\sum_k\frac{1}{\beta\nu(k)}|\vec{\phi}_{\bf k
}|^2+\nonumber\\ &&\sum_R\vec{S'}_{\bf R}\vec{\phi}_{\bf R}+ \beta
D'\sum_{R}(\hat x_{\bf R}\vec{S'}_{\bf R})^2\Big\},
\end{eqnarray}
where $\vec{\phi}_{\bf k}$ is the Fourier transform of the
$m$-component field variable $\vec{\phi}_{\bf R}$, introduced by
the Stratonovich-Hubbard transformation. The integral in
(\ref{represent}) means a functional integration in the space of
the field variables $\vec\phi$.
 Here, we omit the
normalization coefficient before the integral since it does not
alter the behaviour of system near critical point.

Now the trace operation for the spins concerns only the last two
terms of the exponent in (\ref{represent}). Expanding the
exponential function in a Taylor series, performing the
integration with use of generalized $m$-dimensional spherical
coordinates and finally reexponentiating the obtained series, one
gets:
\begin{eqnarray}\label{frepresent}
Z(\{\hat x_{\bf R}\})&\sim& \int (d\vec{\phi})\exp
\Big\{{-}\frac{1}{4}\sum_k\frac{|\vec{\phi}_{\bf k
}|^2}{\beta\nu(k)}{+}\frac{\sum_{\bf R}|\vec{\phi}_{\bf R
}|^2}{2m}\left(1-\frac{2\beta D'}{m(m+2)}\right){-}\nonumber\\
&&\frac{ \sum_{\bf R}|\vec{\phi}_{\bf R}|^4(m+4+\beta
D'(m+6))}{4m^2(m+2)(m+4)}{+}\frac{ \beta D' }{m(m+2)}\sum_{\bf
R}\left(\vec{\phi}_{\bf R}\hat x_{\bf R}\right)^2{-}\nonumber\\
&&\frac{2 \beta D'\sum_{\bf R}|\vec{\phi}_{\bf
R}|^2\left(\vec{\phi}_{\bf R}\hat x_{\bf
R}\right)^2}{m^2(m+2)(m+4)}+\dots\Big\},
\end{eqnarray}
Here, we omit all terms independent on $\phi$ since they give only
a shift of the free energy.
 Expression
(\ref{frepresent}) has been obtained under  assumption of a weak
anisotropy $D'<<1$, thus the only terms in first order in $D'$
were taken into account. Furthermore, there is no interest in
terms of order higher then fourth order in $\phi$, since they are
irrelevant in the RG calculations of the critical properties of
the model. They may be relevant when multicritical behaviour is
studied.

Since the small wave lengths are relevant for the  critical
behaviour, it is possible to use the expansion of the interaction
Fourier image at small $\bf k$: ${(\beta\nu(k))}^{-1}=c_0+b_0{\bf
k}^2+\dots$. Then transforming back to real space and passing to
the continuous limit by replacing summation over $\bf R$ by the
familiar $d$-dimensional integral one obtains:
\begin{eqnarray}\label{zfor1m}
Z(\{\hat x_{\bf R}\}\sim\int(d\vec{\phi}) e^{{\mathcal H}(\hat
x_{\bf R },\vec{\phi}_{\bf R})},
\end{eqnarray}
where
\begin{eqnarray}\label{ffrepresent1}
{\mathcal H}({\hat x_{\bf R},\vec{\phi}_{\bf R }})&=&-\int d^d R
 \Big\{\frac{1}{2}\left[r_0|\vec{\phi}_{\bf R
}|^2+\left|\vec{\nabla}\vec{\phi}_{\bf R }\right|^2\right]{-}D_1
\left(\vec{\phi}_{\bf R}\hat x_{\bf
R}\right)^2{+}v_1|\vec{\phi}_{\bf
R}|^4+\nonumber\\&&z|\vec{\phi}_{\bf R}|^2\left(\vec{\phi}_{\bf
R}\hat x_{\bf R}\right)^2+\dots\Big\}.
\end{eqnarray}
Here, the field $\phi$ was normalized to get coefficient for
gradient term equal to unity and
$r_0=\frac{1{}}{b_0}\left(c_0-\frac{2}{m}(1-\frac{2\beta
D'}{m(m+2)})\right)$, $v_1=\frac{(m+4+\beta
D'(m+6))}{{b_0}^2m^2(m+2)(m+4)}$, $D_1=\frac{4 \beta D'
}{{b_0}^2m(m+2)}$, $z=\frac{8\beta D'}{{b_0}^2m^2(m+2)(m+4)}$.


\begin{thebibliography}{150}
\bibitem{note} Note that the resummation is performed for the
polynomial for the inverse exponent.
\end{thebibliography}
\end{document}